\documentclass{article}%
\usepackage{graphicx}
\usepackage{amsmath}
\usepackage{epic}%
\usepackage{amsfonts}%
\usepackage{amssymb}
\begin{document}

\title{Ursell operators in statistical physics of dense systems: the role of high
order operators and of exchange cycles}
\author{J.N. Fuchs, M.\ Holzmann\thanks{Present address: Physics Department,
University of Illinois at Urbana-Champaign, 1110 W. Green St., Urbana, Il
61801, USA} \thinspace and F.\ Lalo\"{e}\\LKB, D\'{e}partement de Physique de l'ENS\\24 rue Lhomond, F 75005 Paris, France}
\maketitle

\begin{abstract}
The purpose of this article is to discuss cluster expansions in dense quantum
systems as well as their interconnection with exchange cycles.\ We show in
general how the Ursell operators of order $l\geq3$ contribute to an
exponential which corresponds to a mean-field energy involving the second
operator $U_{2}$, instead of the potential itself as usual - in other words,
the mean-field correction is expressed in terms of a modification of a local
Boltzmann equilibrium.\ In a first part, we consider classical statistical
mechanics and recall the relation between the reducible part of the classical
cluster integrals and the mean-field; we introduce an alternative method to
obtain the linear density contribution to the mean-field, which is based on
the notion of tree-diagrams and provides a preview of the subsequent quantum
calculations.\ We then proceed to study quantum particles with Boltzmann
statistics (distinguishable particles) and show that each Ursell operator
$U_{n}$ with $n\geq3$ contains a ``tree-reducible part'', which groups
naturally with $U_{2}$ through a linear chain of binary interactions; this
part contributes to the associated mean-field experienced by particles in the
fluid.\ The irreducible part, on the other hand, corresponds to the effects
associated with three (or more) particles interacting all together at the same
time.\ We then show that the same algebra holds in the case of Fermi or Bose
particles, and discuss physically the role of the exchange cycles, combined
with interactions.\ Bose condensed systems are not considered at this stage.
The similarities and differences between Boltzmann and quantum statistics are
illustrated by this approach, in contrast with field theoretical or Green's
functions methods, which do not allow a separate study of the role of quantum
statistics and dynamics.

\end{abstract}

\section{Introduction}

A widely used formalism in quantum statistical physics is the formalism of
\ Green's functions \cite{Green-1, Green-2, Green-3}, where the techniques of
second quantization and field theory are used from the beginning; the notion
of exchange operators of indistinguishable particles is of course contained,
but in a completely implicit way.\ In the formalism of Ursell operators
\cite{Ursell-1, Ursell-1-bis, Ursell-2}, the starting point is first
quantization with numbered particles, so that the role of exchange cycles
becomes completely explicit: these cycles appear clearly in all diagrams and,
for instance, they are the only source of diagrams for the ideal quantum gas.
This reduces the distance between the formalism and, for instance, numerical
calculations such as the PIMC method (Path Integral Monte Carlo), where
particles are also numbered and the exchange cycles are explicitly sampled by
random choices. Moreover, it becomes possible to assume that the particles
obey Boltzmann statistics, just by ``switching off the cycles'', a task that
would be difficult in the Green's function formalism. Needless to say, this
does not mean that Green's functions are, in general, less powerful than the
Ursell formalism! The opposite is actually closer to reality: for instance,
Green's functions handle time-dependent problems easily, while this is not the
case in the Ursell formalism.\ But it remains true that, if one is interested
in a detailed discussion of the effects of quantum statistics, it becomes more
straightforward to resort to the Ursell formalism.

In this article, we consider dense systems, for which it is not necessarily
possible to limit oneself to first order density effects.\ In contrast to the
situation in a dilute gas, a given particle may interact frequently with
several others at the same time, and even liquefaction may take
place.\ Therefore, we will no longer ignore all Ursell operators beyond
$U_{2}$, as was done in most of previous work in this formalism; operators
$U_{3}$, $U_{4}$, etc.\ now become important. One may actually wonder what the
role of these higher order operators is in general, and why exactly it is
possible to ignore their role in a dilute system, as was done in
\cite{Ursell-2} for instance. We will see that part of their contribution
(what we will call their tree reducible part) groups naturally with $U_{2}$
through a chain of binary interactions, and builds an exponential of the
mean-field energy.\ In other words, instead of making the problem more
complicated, this contribution of the higher order operators builds exactly
the exponential of $U_{2}$ that is needed to reconstruct a simple and natural
expression of the mean-field.\ Nevertheless, this mean-field is expressed in
terms of the matrix elements of $U_{2}$, instead of the usual matrix elements
of the potential itself; in a sense, what we obtain is the exponential of an
exponential, since $U_{2}$ itself contains exponentials of the Hamiltonian and
corresponds physically to the local change of the Boltzmann equilibrium.\ As a
consequence, the logarithms that appeared in \cite{Ursell-2}, and had to be
expanded to first order in density, are actually spurious - in other words,
this first order expansion was actually unnecessary.\ In addition, the rest of
the contribution of the higher order Ursell operators, the irreducible part,
vanishes unless three particles (or more) are all close together, and is
really characteristic of many-body collisions and of dense systems.

\section{Classical statistical physics}

\label{classical}

In this part, we quickly review the classical cluster expansion for the
parametric equation of state of a fluid; we first recall the results of
classical statistical mechanics as a point of comparison.\ 

\subsection{General formalism}

The general expression of the equation of state was derived by Mayer
\cite{Mayer} and Ursell \cite{Ursell original}; see also the books by
Uhlenbeck and Ford \cite{Uhlenbeck} and Hansen and McDonald \cite{Hansen}. A
classical system of \ massive particles, with mass $m$, is supposed to be
contained in a box of finite volume $\mathcal{V}$, with periodic boundary
conditions (translationally invariant system); their Hamiltonian is the sum of
the kinetic energies plus the interaction energy, which is the sum over all
pairs of particles of the interparticle pair potential $V_{ij}=V(\mathbf{r_{i}%
}-\mathbf{r_{j}})$. An useful function in the calculation is the Mayer
function $f_{ij}$ defined by:
\begin{equation}
f_{ij}\equiv\exp(-\beta V_{ij})-1 \label{6}%
\end{equation}
This function is everywhere bounded and goes to almost zero when the distance
between particles $i$ and $j$ is much larger than the range of the pair
potential; it is the classical equivalent of a second Ursell operator
$\overline{U}_{2}(i,j)$ \cite{Ursell-2}.

The classical cluster expansion is the expansion, in the grand canonical
ensemble, of the pressure $p$ and of the density $\rho\ $of the gas in series
of the fugacity $z=e^{\beta\mu}$, where $\mu$ is the chemical potential and
$\beta=1/k_{B}T$ \ is the inverse temperature. If $\lambda$ is the de Broglie
thermal wavelength, the expansion can be written as:
\begin{equation}
\beta p=\frac{\ln Z_{gc}}{\mathcal{V}}=\sum_{l=1}^{\infty}b_{l}(T,\mathcal{V}%
)\left(  \frac{z}{\lambda^{3}}\right)  ^{l} \label{1}%
\end{equation}
with:
\begin{equation}
\rho=\frac{\langle N\rangle}{\mathcal{V}}=z\frac{\partial}{\partial z}\left(
\frac{\ln Z_{gc}}{\mathcal{V}}\right)  =\sum_{l=1}^{\infty}l\ b_{l}%
(T,\mathcal{V})\left(  \frac{z}{\lambda^{3}}\right)  ^{l} \label{2}%
\end{equation}
where $Z_{gc}(\beta,\mathcal{V},z)$ is the grand canonical partition function;
the definition of the cluster integrals $b_{l}(T,\mathcal{V})$ is the same as
that in the book of Mayer \& Mayer \cite{Mayer}. The first coefficients are
given by:
\begin{equation}
b_{1}(T,\mathcal{V})=\frac{1}{1!\mathcal{V}}\int d^{3}r_{1}=1 \label{3}%
\end{equation}%
\begin{equation}
b_{2}(T,\mathcal{V})=\frac{1}{2!\mathcal{V}}\int d^{3}r_{1}\,d^{3}%
r_{2}\,\,f_{12}=\frac{1}{2!}\int d^{3}r_{12}\,\,f(r_{12}) \label{4}%
\end{equation}%
\begin{equation}
b_{3}(T,\mathcal{V})=\frac{1}{3!\mathcal{V}}\int d^{3}r_{1}\,d^{3}r_{2}%
\,d^{3}r_{3}[f_{12}f_{13}+f_{12}f_{23}+f_{13}f_{23}+f_{12}f_{13}f_{23}]
\label{5}%
\end{equation}
and so on for $l\geq4$; these definitions include a $1/\mathcal{V}$ factor so
that $b_{l}(T,\mathcal{V})$ remains intensive when the volume tends to infinity.

It is useful to introduce diagrams to represent the integrals. A $l$-particle
cluster diagram is made of \ $l$ numbered circles, representing the particles,
between which one draws any number of lines (also called links, and
representing the $f_{ij}$'s), each line joining distinct pairs of circles. The
diagram is said to be connected when one can go from any particle to any other
particle in the cluster following the lines. The general definition of the
cluster integrals is then:%

\begin{equation}
b_{l}(T,\mathcal{V})\equiv\frac{1}{l!\,\mathcal{V}}\times\text{sum over all
possible }l\text{-particle clusters} \label{5bis}%
\end{equation}
For example:
\begin{equation}
b_{1}(T,\mathcal{V})=\frac{1}{1!\mathcal{V}}\left(
\begin{picture}
(10,0) \put(5,3){\circle*{4}}
\end{picture}
\right)  =1 \label{7}%
\end{equation}
\label{7quart}%
\begin{equation}
b_{2}(T,\mathcal{V})=\frac{1}{2!\mathcal{V}}\left(
\begin{picture}
(20,0) \put(2,3){\circle*{4}} \put(2,3){\line(1,0){16}} \put(18,3){\circle
*{4}}
\end{picture}
\right)  \label{7bis}%
\end{equation}
and:
\begin{equation}
b_{3}(T,\mathcal{V})=\frac{1}{3!\mathcal{V}}\left(
\begin{picture}
(24,15) \put(11,13){\circle*{4}}\put(2,-2){\circle*{4}} \put(20,-2){\circle
*{4}} \drawline(2,-2)(20,-2) \drawline(20,-2)(11,13)
\end{picture}
+%
\begin{picture}
(24,15) \put(11,13){\circle*{4}} \put(2,-2){\circle*{4}}\put(20,-2){\circle
*{4}} \drawline(11,13)(2,-2) \drawline(20,-2)(11,13)
\end{picture}
+%
\begin{picture}
(24,15) \put(11,13){\circle*{4}}\put(2,-2){\circle*{4}}\put(20,-2){\circle
*{4}} \drawline(11,13)(2,-2) \drawline(2,-2)(20,-2)
\end{picture}
+%
\begin{picture}
(24,15) \put(11,13){\circle*{4}} \put(2,-2){\circle*{4}} \put(20,-2){\circle
*{4}} \drawline(11,13)(2,-2) \drawline(2,-2)(20,-2) \drawline(20,-2)(11,13)
\end{picture}
\right)  \label{7ter}%
\end{equation}

\subsection{Irreducible diagrams, energy shift and equation of state}

It turns out that, defined in this way, diagrams carry redundant information:
as soon as $l\geq3$, most of the diagrams contributing to $b_{l}$ can be
constructed as products of smaller diagrams (already contained in
$b_{l^{^{\prime}}}$ with $l^{\prime}<l$). For instance, this is the case of
all diagrams in the right hand side of (\ref{7ter}), except the last.\ It then
becomes convenient to define the notion of irreducibility: an irreducible
cluster is such that, if any line (which is essentially a $b_{2}$) is removed
from it, it never splits into disconnected clusters; conversely, in a
reducible cluster, it is possible to find at least one line that, when cut,
will decompose the result into two separate clusters (since each line is
essentially a $b_{2}$, reducibility and irreducibility are defined here in
terms of binary coefficients $b_{2}$'s, but more general definitions are conceivable).

Irreducible diagrams play a special role if one introduces an exponentiation
of the fugacity expansion of the density (\ref{2}). Let us define an energy
shift $\Delta$, which shifts all energy levels $e_{k}$, by the relation:
\begin{equation}
\rho=e^{-\beta\left(  \Delta-\mu\right)  }\frac{1}{\lambda^{3}} \label{8}%
\end{equation}
It can then be shown (see \cite{Mayer bis}, \cite{Hansen bis}, \cite{Morita},
\cite{Dominicis}) that $\Delta$ is given by the series:
\begin{equation}
-\beta\Delta=\sum_{l=1}^{\infty}\beta_{l}(T,\mathcal{V})\,\rho^{l} \label{9}%
\end{equation}
where the new coefficients $\beta_{l}$ are defined by:
\begin{equation}
\beta_{l-1}(T,\mathcal{V})\equiv\frac{1}{(l-1)!\mathcal{V}}\,\times\text{sum
over all irreducible }l\text{-particle clusters} \label{11}%
\end{equation}
or, equivalently:
\begin{equation}
\beta_{l-1}(T,\mathcal{V})\equiv l\,\times\text{Irreducible part of }%
b_{l}(T,\mathcal{V}) \label{12}%
\end{equation}
For example:
\begin{equation}
\beta_{1}(T,\mathcal{V})=\frac{1}{\mathcal{V}}\left(
\begin{picture}
(20,0) \put(2,3){\circle*{4}} \put(2,3){\line(1,0){16}} \put(18,3){\circle
*{4}}
\end{picture}
\right)  =\int d^{3}r_{12}\,\,f(r_{12}) \label{12bis}%
\end{equation}
and:%
\begin{equation}
\beta_{2}(T,\mathcal{V})=\frac{1}{2!\mathcal{V}}\left(
\begin{picture}
(24,15) \put(11,13){\circle*{4}} \put(2,-2){\circle*{4}} \put(20,-2){\circle
*{4}} \drawline(11,13)(2,-2) \drawline(2,-2)(20,-2) \drawline(20,-2)(11,13)
\end{picture}
\right)  \label{12ter}%
\end{equation}
One can also show that irreducible clusters are directly related to the
equation of state (at finite volume), from which the variable $z$ has been
eliminated:
\begin{equation}
\frac{\beta p}{\rho}=1-\sum_{l=1}^{\infty}\frac{l}{l+1}\,\beta_{l}%
(T,\mathcal{V})\,\,\rho^{l} \label{13}%
\end{equation}
The proof of this result is given in the book by Mayer \& Mayer \cite{Mayer}%
.\ Another method of calculation was introduced by Van Kampen
\cite{Van-Kampen}, who works with the canonical ensemble instead of grand
canonical; see also \cite{Mullin}.\ In these calculations, the thermodynamical
limit is taken before the $l$ summation while, here, we do not take this
limit; the reason is that, as discussed by Lee and Yang \cite{Lee}, taking
directly the limit for each coefficient reduces the validity of the
calculation to the gaseous phase only.

\subsection{Tree-reducible diagrams; mean-field}

\label{classical-tree}

\begin{figure}
[ptb]
\begin{center}
\includegraphics[
height=1.535in,
width=1.7167in
]%
{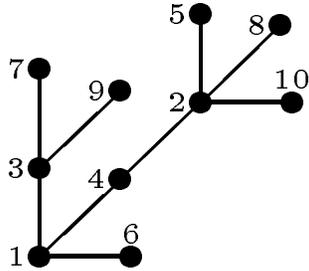}%
\caption{A tree-diagram corresponding to a 10-particle cluster; the tree
symbolizes one of the terms that are contained in the tree-reducible part of
$b_{10}$. The construction method that is used to build the branches from the
structure of the integral is explained in the text.}%
\label{figure1}%
\end{center}
\end{figure}

Let us now suppose that, among all clusters contributing to $b_{l}$, we keep
only those having $l$ particles and $l-1$ links.\ They correspond to the
minimally connected diagrams: if any link is removed, these diagrams split
into two different diagrams.\ They are, not only obviously reducible (as soon
as $l\geq3$), but also ``fully reducible'', since the removal of any line
splits the diagram into disconnected clusters. For instance, the fully
reducible part of $b_{3}$ is:%
\begin{equation}
b_{3}^{R}(T)=\frac{1}{3!\mathcal{V}}\int d^{3}r_{1}\,d^{3}r_{2}\,d^{3}%
r_{3}[f_{12}f_{13}+f_{12}f_{23}+f_{13}f_{23}] \label{16}%
\end{equation}
from which the last term that appear on the right hand side of (\ref{5}) has
been eliminated. Graphically, we will represent these clusters by diagrams
that have the structure of a tree, such as that of figure \ref{figure1}, and
for this reason we shall call them ``tree-reducible diagrams''.

In order to construct this tree-diagram, we have to choose one numbered
particle as the ``root particle'', for instance particle $1$. Then, among all
particles $j$ that are connected to particle $1$ by a link $f_{1j}$, we select
that with the smallest value of $j$; let us call $k$ this particle.\ The same
operation is then made again from particle $k$: one identifies the particle
$m$ to which it is related which has the smallest numbering, and puts this
particle as the next in the branch; the same operation goes on until one
reaches the end of the first branch, which is drawn vertically by
convention.\ One then goes back along this branch towards the origin and
identifies the first particle which is connected to another particle outside
of the branch; this particle is a branching point, the source of another
branch, which is built in exactly the same way, and drawn directly on the
right of the first branch.\ The same process continues and one adds successive
branches, which are drawn in a clockwise order from their source; it stops
when all particles are included in the tree.\ This construction provides a
well defined geometry for the tree-diagram but, clearly, different
tree-diagrams may have the same numerical value\footnote{In fact, the value of
these diagrams depends only on the total number of nodes (i.e. particles) $l$
(see Appendix I).}.\ It is nevertheless convenient for our reasoning below to
classify their contributions according to the various structures of trees.%

%TCIMACRO{\FRAME{ftbpFU}{1.7167in}{1.535in}{0pt}{\Qcb{A tree-diagram
%corresponding to a 10-particle cluster; the tree symbolizes one of the terms
%that are contained in the tree-reducible part of $b_{10}$. The construction
%method that is used to build the branches from the structure of the integral
%is explained in the text.}}{\Qlb{figure1}}{figure1.eps}%
%width 1.7167in;  height 1.535in;  depth 0pt;  original-width 1.0274in;
%original-height 0.915in;  cropleft "0";  croptop "1";  cropright "1";
%cropbottom "0";  filename 'figure1.eps';file-properties "XNPEU";}}}%
%BeginExpansion
%EndExpansion
For instance, the reducible part of $b_{3}$ then becomes:%
\begin{equation}
b_{3}^{R}(T,\mathcal{V})=\frac{1}{3!}(2\times%
\begin{picture}
(10,30) \put(2,2){\circle*{4}} \put(2,16){\circle*{4}}\put(2,30){\circle*{4}%
}\drawline(2,2)(2,16) \drawline(2,16)(2,30)
\end{picture}
+%
\begin{picture}
(20,20) \put(2,2){\circle*{4}} \put(2,16){\circle*{4}} \put(16,2){\circle*{4}%
}\drawline(2,4)(2,16) \drawline(4,2)(16,2)
\end{picture}
) \label{17}%
\end{equation}
(from now on, we assume that the factor $1/\mathcal{V}$ is included in the
value of the tree-diagram, in order to simplify the notation).

\ Let us now consider a given tree-diagram and evaluate its weight; the
question is to determine how many different numberings are compatible with the
rules that we have used to build it. Suppose then that we put random numbers
into the $l-1$ nodes that are available in the tree.\ The result will be
acceptable only if a correct clockwise ordering of the particle numbers is
obtained at each bifurcation node.\ Let us call $r_{1}$, $r_{2}$, .. the
branching factors (or ramification factors) of the nodes, i.e. the number of
secondary branches at each node; a linear diagram has only $r=1$, a diagram
with one binary bifurcation has one $r=2$, and $r$ takes the values $3$, $4$,
etc. when more branches start from the same source. Each node will then
introduce a probability $1/r_{i}!$ \ for this correct ordering to be
obtained.\ Since there are $(l-1)!$ ways of distributing the $l-1$ particles
among the nodes, the final result is that the weight of the tree-diagram is:%
\begin{equation}
\frac{(l-1)!}{\prod_{i}r_{i}!} \label{18}%
\end{equation}
If we call $T_{diag}$ the value of a tree-diagram (including the
$1/\mathcal{V}$ factor), we can then express the tree-reducible $l$ particles
cluster as:%
\begin{equation}
l!\ b_{l}^{R}(T)=\sum_{\left\{  tree\,\,diagrams\right\}  }\frac{(l-1)!}%
{\Pi_{i}(r_{i})!}\,T_{diag.} \label{19}%
\end{equation}
Inserting this expression into the $z$-expansion of the density (\ref{2}), we
notice that various simplifications take place in the coefficients, so that we
obtain:%
\begin{equation}%
\begin{split}
\rho &  =\frac{z}{\lambda^{3}}\left\{  1+\frac{z/\lambda^{3}}{1!}\times%
\begin{picture}
(5,20) \put(2,2){\circle*{4}} \put(2,16){\circle*{4}}\drawline(2,4)(2,16)
\end{picture}
+\frac{(z/\lambda^{3})^{2}}{2!}\left(  \frac{2!}{2!}\times%
\begin{picture}
(25,20) \put(2,2){\circle*{4}}\put(2,16){\circle*{4}}\put(16,2){\circle*{4}}
\drawline(2,4)(2,16) \drawline(4,2)(16,2)
\end{picture}
+\frac{2!}{1!}\times%
\begin{picture}
(5,40) \put(2,2){\circle*{4}} \put(2,16){\circle*{4}} \put(2,30){\circle*{4}%
}\drawline(2,4)(2,16) \drawline(2,16)(2,30)
\end{picture}
\right)  +...\right. \\
&  \left.  +\frac{(z/\lambda^{3})^{l-1}}{(l-1)!}\times\frac{(l-1)!}{\Pi
_{i}(r_{i})!}\times%
\begin{picture}
(40,50) \put(2,2){\circle*{4}}\put(2,18){\circle*{4}}\put(2,36){\circle*{4}}
\put(18,2){\circle*{4}}\put(16,16){\circle*{4}}\put(16,32){\circle*{4}}
\put(30,30){\circle*{4}}\put(44,44){\circle*{4}}\put(30,46){\circle*{4}}
\put(46,30){\circle*{4}}\drawline(2,4)(2,18) \drawline(4,2)(18,2)
\drawline(2,18)(2,36) \drawline(3,3)(16,16) \drawline(16,16)(30,30)
\drawline(2,18)(16,32) \drawline(30,30)(44,44) \drawline(30,30)(30,46)
\drawline(30,30)(46,30)
\end{picture}
+...\right\}
\end{split}
\label{20}%
\end{equation}
It is now possible to regroup diagrams according to the value of the branching
factor $r_{1}$ at their root:
\begin{equation}%
\begin{array}
[c]{ll}%
\rho= & \frac{z}{\lambda^{3}}\left\{  1+\frac{1}{1!}\left[  \frac{z}%
{\lambda^{3}}%
\begin{picture}
(5,20) \put(2,2){\circle*{4}} \put(2,16){\circle*{4}}\drawline(2,4)(2,16)
\end{picture}
+\left(  \frac{z}{\lambda^{3}}\right)  ^{2}%
\begin{picture}
(5,40) \put(2,2){\circle*{4}} \put(2,16){\circle*{4}}\put(2,30){\circle*{4}}
\drawline(2,4)(2,16) \drawline(2,16)(2,30)
\end{picture}
+\left(  \frac{z}{\lambda^{3}}\right)  ^{3}%
\begin{picture}
(20,40) \put(2,2){\circle*{4}} \put(2,16){\circle*{4}}\put(16,16){\circle*{4}%
}\put(2,30){\circle*{4}} \drawline(4,16)(16,16) \drawline(2,4)(2,16)
\drawline(2,16)(2,30)
\end{picture}
+...\right]  \right. \\
& \left.  +\frac{1}{2!}\left[  \left(  \frac{z}{\lambda^{3}}\right)  ^{2}%
\begin{picture}
(25,20) \put(2,2){\circle*{4}}\put(2,16){\circle*{4}}\put(16,2){\circle*{4}}
\drawline(2,4)(2,16) \drawline(4,2)(16,2)
\end{picture}
+...\right]  +...\right\}
\end{array}
\label{21}%
\end{equation}
One then notices a sort of self-similarity property of the expansion: at each
secondary nodes, one gets an expansion that provides again all the terms that
are contained in $\rho$.\ Therefore:
\begin{equation}
\rho=\frac{z}{\lambda^{3}}\left\{  1+%
\begin{picture}
(5,20) \put(2,2){\circle*{4}}\put(0,16){\frame{$\scriptstyle{\rho}$}}
\drawline(2,4)(2,16)
\end{picture}
+\frac{1}{2!}\times%
\begin{picture}
(25,20) \put(2,2){\circle*{4}}\put(0,16){\frame{$\scriptstyle{\rho}$}}%
\put(16,0){\frame{$\scriptstyle{\rho}$}} \drawline(2,4)(2,16) \drawline
(4,2)(16,2)
\end{picture}
+...+\frac{1}{r_{1}!}\times%
\begin{picture}
(25,20) \put(2,2){\circle*{4}} \put(0,16){\frame{$\scriptstyle{\rho}$}}%
\put(16,0){\frame{$\scriptstyle{\rho}$}}\put(14,14){\frame{$\scriptstyle{\rho
}$}} \drawline(2,4)(2,16) \drawline(4,2)(16,2) \drawline(3,3)(14,14)
\end{picture}
+...\right\}  \label{22}%
\end{equation}
We now recognize the development of an exponential:%
\[
\rho=\frac{z}{\lambda^{3}}\left\{  1+\beta_{1}\rho+\frac{1}{2!}\times\left[
\beta_{1}\rho\right]  ^{2}+...+\frac{1}{r_{1}!}\times\left[  \beta_{1}%
\rho\right]  ^{r_{1}}+...\right\}
\]
where $\beta_{1}=2b_{2}$ is defined in (\ref{12bis}), and finally obtain the
very simple result:
\begin{equation}
\rho=\frac{z}{\lambda^{3}}\exp{(}\beta_{1}\rho{)} \label{14}%
\end{equation}
or:
\begin{equation}
-\beta\Delta=\beta_{1}\rho\label{13ter}%
\end{equation}
Therefore, when only tree-reducible diagrams are taken into account, the
energy shift becomes exactly proportional to the density; we recover what is
usually called the mean-field approximation.\ The method we have used is
different from the method of refs.\ \cite{Mayer bis, Hansen bis, Morita,
Dominicis}; it relies on the branching properties of tree-reducible diagrams
and not on the use of complex variables.\ Moreover, as we will see below, the
method can be transposed to quantum mechanics.

Of course, equation (\ref{13ter}) can also be obtained by the same method as
these references, in a particular case: one assumes that the only non-zero
irreducible cluster integral is $\beta_{1}$.\ Clearly, in this situation, all
cluster integrals reduce to their tree-reducible value $b_{l}^{R}$, which can
be expressed as a product of terms $b_{2}=$ $\beta_{1}/2$ (see Appendix I):
\begin{equation}
b_{l}^{R}\equiv\frac{1}{l!\ }N_{l}\,\,\beta_{1}^{l-1} \label{13bis}%
\end{equation}
where $N_{l}$ is the number of clusters with $l$ numbered particles and $l-1$
links, i.e. the number of classical tree-diagrams containing $l$ particles:%
\begin{equation}
N_{l}=\sum_{\left\{  tree\,\,diagrams\right\}  }\frac{(l-1)!}{\Pi_{i}(r_{i})!}
\label{13quater}%
\end{equation}
If we insert the corresponding value of $b_{l}$ into equations (\ref{1}) and
(\ref{2}), we obtain again equation (\ref{13ter}).\ Moreover, equation
(\ref{13}) then becomes:
\begin{equation}
\frac{\beta p}{\rho}=1-\frac{1}{2}\beta_{1}(T,\mathcal{V})\,\,\rho
=1-b_{2}(T,\mathcal{V})\,\,\rho\label{15}%
\end{equation}
which contains only a linear density correction.

\section{Quantum statistical physics: Boltzmann statistics}

\label{quantum}

We now leave classical statistical mechanics and reason within quantum
mechanics; in a first step, we consider distinguishable particles (Boltzmann
particles), postponing the discussion of the effects of quantum statistics to
the next section (bosons or fermions). We first introduce our notation and
then discuss the introduction of exponentials from the structure of tree diagrams.

\subsection{Notation}

\label{notation}

We assume that the Hamiltonian of the system of $N$ non-relativistic particles
is:
\begin{equation}
H=\sum_{i=1}^{N}H_{0}(i)+\sum_{i<j}V_{ij} \label{a1}%
\end{equation}
where $H_{0}(i)$ is the one-particle energy (sum of its kinetic energy plus
coupling to an external potential) and where the binary interaction potential
$V_{ij}$ is a function of the distance between particles $i$ and $j$:
\begin{equation}
V_{ij}=V(\left|  \mathbf{r}_{i}-\mathbf{r}_{j}\right|  )=V(r_{ij}) \label{a2}%
\end{equation}
As in \cite{Ursell-1}, we define the Ursell operators:
\begin{equation}
U_{1}(1)=\exp\left[  -\beta\,H_{0}(1)\right]  \label{a3}%
\end{equation}
and:
\begin{equation}
U_{2}(1,2)=\exp\left[  -\beta\left[  H_{0}(1)+H_{0}(2)+V_{12}\right]  \right]
-U_{1}(1)U_{1}(2) \label{a4}%
\end{equation}
and so on for higher order Ursell operators $U_{3}(1,2,3)$, $U_{4}(1,2,3,4)$,
etc. All the $U_{l}$'s for $l\geq2$ have a clustering property; for instance,
the diagonal matrix elements of $<\mathbf{r}_{1},\mathbf{r}_{2}\mid U_{2}%
\mid\mathbf{r}_{1},\mathbf{r}_{2}>$ tend towards zero when the distance
$\mid\mathbf{r}_{1}-\mathbf{r}_{2}\mid$ becomes larger than some microscopic
distance; in other words, the matrix elements of the same operator in the
momentum representation are proportional to the inverse volume $1/\mathcal{V}%
$.\ One can then express the canonical partition function $Z_{N}$ as a sum of
products of traces of operators $U_{l}$'s, each corresponding to a given
diagram.\ Here, since we are dealing with Boltzmann particles, the number of
diagrams is smaller than in \cite{Ursell-1}; exchange cycles do not occur,
which is equivalent to limit them to cycles containing one particle only - in
other words we exclude all those diagrams that contain horizontal
lines\footnote{As usual, we retain the overall $1/N!$ factor in the
symmetrizer, in order to avoid some well-known difficulties of classical
statistical mechanics (Gibbs paradox, etc.).}.\ Going to the grand canonical
ensemble, one can then show that the logarithm of the corresponding partition
function $Z_{g.c.}$ is given by the following sum of traces:
\begin{equation}
\ln Z_{g.c.}=\sum_{l=1}^{\infty}\frac{z^{l}}{l!}Tr_{1,2,...,l}\left\{
U_{l}(1,2,...,l)\right\}  \label{a5}%
\end{equation}
where the factor $1/l!$ corresponds to the weight of the diagram, which arises
because there are $l!$ equivalent ways to distribute $l$ numbered particles
inside the Ursell operator, as discussed in \cite{Ursell-1}.

Rather than studying a thermodynamic potential, it is often more convenient to
focus the discussion on the single particle density operator $\rho_{I}(1)$, as
in ref.\ \cite{Ursell-2}.\ One then gets:
\begin{equation}
\rho_{I}(1)=zU_{1}(1)+z^{2}\,Tr_{2}\left\{  U_{2}(1,2)\right\}  +\frac{z^{3}%
}{2!}\,Tr_{2,3}\left\{  U_{3}(1,2,3)\right\}  +... \label{a6}%
\end{equation}
Now, because particle $1$ is ``tagged'', it plays a special role, so that the
generic term of this series is:
\begin{equation}
\frac{z^{l}}{(l-1)!}Tr_{2,3,.....l}\left\{  U_{l}(1,2,3,...,l)\right\}
\label{a7}%
\end{equation}
with a weight $1/(l-1)!$ corresponding to the equivalent distributions of all
the untagged particles.\ In the usual graphical representation or Ursell
diagrams \cite{Ursell-1,Ursell-2}, where vertical lines symbolize $U_{l}$
operators (for $l\geq2$) and horizontal lines exchange cycles\ (we have
already mentioned that, since here we are dealing with distinguishable
particles, no cycle of length greater than one occurs), \ equation (\ref{a6})
becomes:%
\begin{equation}
\rho_{I}(1)=%
\begin{picture}
(20,20) \put(2,0){\line(0,1){4}} \put(14,0){\line(0,1){4}} \put(2,2){\line
(1,0){12}}
\end{picture}
+%
\begin{picture}
(20,20) \put(2,0){\line(0,1){4}} \put(14,0){\line(0,1){4}} \put(2,2){\line
(1,0){12}} \put(2,12){\line(0,1){4}} \put(14,12){\line(0,1){4}} \put
(2,14){\line(1,0){12}} \put(7,2){\line(0,1){12}} \put(9,2){\line(0,1){12}}%
\end{picture}
+\frac{1}{2!}\times%
\begin{picture}
(20,20) \put(2,0){\line(0,1){4}} \put(14,0){\line(0,1){4}} \put(2,2){\line
(1,0){12}} \put(2,12){\line(0,1){4}} \put(14,12){\line(0,1){4}} \put
(2,14){\line(1,0){12}}\put(2,24){\line(0,1){4}} \put(14,24){\line(0,1){4}}
\put(2,26){\line(1,0){12}} \put(6,2){\line(0,1){24}} \put(8,2){\line(0,1){24}}
\put(10,2){\line(0,1){24}}
\end{picture}
+... \label{a6bis}%
\end{equation}
At this stage, it is convenient \cite{Ursell-2,Markus} to express the second
rank operators $U_{2}$ in terms of the operator $\overline{U}_{2}$ defined
by:
\begin{equation}
U_{2}(1,2)=\sqrt{U_{1}(1)U_{1}(2)}\,\,\overline{U}_{2}(1,2)\,\,\sqrt
{U_{1}(1)U_{1}(2)} \label{a8}%
\end{equation}
(for the moment, when $l>2$, we do not specify the exact relation between an
$U_{l}$ operator and $\overline{U}_{l}$, but we will come back to this point
later).\ We note that, here, $\overline{U}_{2}$ is defined symmetrically, with
square roots of $U_{1}$ operators on each side ($U_{1}$ \ is a positive
operator), so that $\overline{U}_{2}$ is Hermitian - this was not the case in
\cite{Ursell-1-ter}.\ The introduction of $\overline{U}_{2}$ has two
advantages.\ First, factorizing the kinetic energy brings the formalism closer
to classical statistical mechanics, where kinetic energy always factorizes out
exactly; in other words, $\overline{U}_{2}(i,j)$ is the quantum equivalent of
$f_{ij}$,\ although it is not strictly analogous (in quantum mechanics
operators do not necessarily commute, so that the kinetic energy can still
play a role in $\overline{U}_{2}(i,j)$).\ Second, in the limit of low
energies, it \ is possible to make use of the MIME (momentum independent
matrix elements) approximation, where the diagonal matrix elements of
$\overline{U}_{2}$ are constants - see for instance the characterization of
the diagonal elements of $\overline{U}_{2}$ in terms of the Ursell length in
\cite{Ursell-1-ter}. Graphically, in order to distinguish $U_{2}$'s from
$\overline{U}_{2}$'s in diagrams, we use two vertical parallel lines in the
former case, as in \cite{Ursell-1} and \cite{Ursell-1-bis}, but only a single
line for $\overline{U}_{2}$'s.

\subsection{Tree-reducible part of $\overline{U}_{l}$ and $U_{l}$}

\label{tree}

For any value of $l$, we now define the tree-reducible operator (or fully
reducible operator) $\overline{U}_{l}^{R}$ by analogy with the tree-reducible
part of a classical cluster $b_{l}$.\ Since the classical links correspond to
$\overline{U}_{2}$ operators in quantum mechanics, and since the minimal
number of links is $l-1$, we will express the fully reducible operator
$\overline{U}_{l}$ as a product of $l-1$ operators $\overline{U}_{2}$; in
addition, we also have to take into account the fact that operators do not
necessarily commute with each other and apply an appropriate
symmetrization.\ We will therefore define $\overline{U}_{l}^{R}$ by a double
sum:
\begin{equation}
\overline{U}_{l}^{R}(1,2,...,l)=\sum_{\left\{  \overline{U}_{2}\right\}
}\,\frac{1}{(l-1)!}\sum_{\left\{  op.\;orderings\right\}  }\overline{U}%
_{2}(.,.)\,\overline{U}_{2}(.,.)\ ...\ \overline{U}_{2}(.,.) \label{a17}%
\end{equation}
The first sum symbolizes all different ways to choose sets of $l-1$ operators
$\overline{U}_{2}$ so that the product correspond to a minimally connected
cluster - in other words, in all classical tree-diagrams, we replace each
$f_{ij}$ by the corresponding $\overline{U}_{2}(i,j)$. The second sum,
together with the factor $1/(l-1)!$, corresponds to an average of the product
of all operators over all possible orderings of these $l-1$ operators. With
this definition, $\overline{U}_{l}^{R}$ is obviously Hermitian as well as
symmetrical with respect to all particles $1$, $2$,..., $l$. For instance:
\begin{equation}%
\begin{array}
[c]{cc}%
\overline{U}_{3}^{R}(1,2,3)= & \frac{1}{2!}[\overline{U}_{2}(1,2)\overline
{U}_{2}(1,3)+\overline{U}_{2}(1,3)\overline{U}_{2}(1,2)]\\
& +\frac{1}{2!}[\overline{U}_{2}(1,2)\overline{U}_{2}(2,3)+\overline{U}%
_{2}(2,3)\overline{U}_{2}(1,2)]\\
& +\frac{1}{2!}[\overline{U}_{2}(1,3)\overline{U}_{2}(2,3)+\overline{U}%
_{2}(2,3)\overline{U}_{2}(1,3)]
\end{array}
\label{a17bis}%
\end{equation}

\begin{figure}
[ptb]
\begin{center}
\includegraphics[
height=1.4641in,
width=2.1819in
]%
{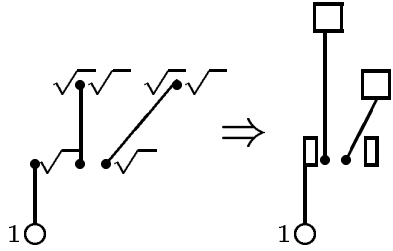}%
\caption{In this diagram, the single lines symbolize $\overline{U}_{2}$
operators (as opposed to double lines which symbolize $U_{2}$'s in Ursell
diagrams).\ The figure illustrates symbolically how $\sqrt{U_{1}}$'s
(symbolised as squareroots $\sqrt{\ }$ in the left part, as rectangles in the
right part) are inserted into each product of $\overline{U}_{2}$'s that
appears in $\overline{U}_{l}^{R}$, in order to build an operator $U_{l}^{R}$.
}%
\label{figure2}%
\end{center}
\end{figure}

We now wish to define an operator $U_{l}^{R}$ that we will call the
tree-reducible part of $U_{l}$. The first idea that comes to mind is to mimic
(\ref{a8}) and to define $U_{l}^{R}$ by just multiplying $\overline{U}_{l}%
^{R}$ on both sides by a product of $\sqrt{U_{1}}$'s.\ This is possible, but
it turns out that this definition would be less convenient than another
slightly different possibility, where some square root operators are inserted
at different places.\ To introduce this definition, we first remark that each
term in the first summation of (\ref{a17}) can be associated with a tree,
exactly as in the classical tree of figure \ref{figure1}, where particle $1$
is put at the root, and all the rest of the diagram is built exactly in the
same way (the $f_{ij}$'s are replaced by $\overline{U}_{2}(i,j)$'s). The
problem now is how to properly ``dress'' this diagram with $\sqrt{U_{1}}$'s in
order to build a proper $U_{l}^{R}$; we will do this operation progressively,
starting from the root.\ For particle $1$, we choose to put $\sqrt{U_{1}(1)}$
on each side of $\overline{U}_{l}^{R}$, as in (\ref{a8}).\ We then progress
along the branches of the tree; we first consider all $\overline{U}_{2}%
(1,j)$'s that start from the root, and associate to each of them one
$\sqrt{U_{1}(j)}$ that is inserted directly on the right side of this
$\overline{U}_{2}$ operator; the other $\sqrt{U_{1}(j)}$ is merely put at the
end, after all the $\overline{U}$'s.\ We then proceed to apply exactly the
same method to the ``second generation'' of $\overline{U}_{2}(j,k)$'s, and
again insert one $\sqrt{U_{1}(k)}$'s directly on their right side, another at
the end, etc..\ We continue this operation until the whole tree is dressed
with $U_{1}$'s; the construction is sketched in figure \ref{figure2}, which
shows the association between the additional $U_{1}$'s and the initial
$\overline{U}_{2}$'s.\ Applying this dressing procedure to each term of the
right hand side of (\ref{a17}) leads to an operator that we note $\widehat
{U}_{l}^{R}$.\ For instance, if $l=3$, we have:%
\begin{equation}%
\begin{array}
[c]{cl}%
\widehat{U}_{3}^{R}(1,2,3)= & \frac{1}{2}\sqrt{U_{1}(1)}\left[  \,\overline
{U}_{2}(1,2)\sqrt{U_{1}(2)}\,\overline{U}_{2}(1,3)\sqrt{U_{1}(3)}+\right. \\
& +\,\overline{U}_{2}(1,3)\sqrt{U_{1}(3)}\,\,\overline{U}_{2}(1,2)\sqrt
{U_{1}(2)}+\\
& \left.  +\,\overline{U}_{2}(1,2)\sqrt{U_{1}(2)}\,\overline{U}_{2}%
(2,3)\sqrt{U_{1}(3)}+....\right]  \,\sqrt{U_{1}(1)U_{1}(2)U_{1}(3)}%
\end{array}
\label{a17ter}%
\end{equation}%

%TCIMACRO{\FRAME{ftbpFU}{2.1819in}{1.4641in}{0pt}{\Qcb{In this diagram, the
%single lines symbolize $\overline{U}_{2}$ operators (as opposed to double
%lines which symbolize $U_{2}$'s in Ursell diagrams).\ The figure illustrates
%symbolically how $\sqrt{U_{1}}$'s (symbolised as squareroots $\sqrt{\ }$ in
%the left part, as rectangles in the right part) are inserted into each product
%of $\overline{U}_{2}$'s that appears in $\overline{U}_{l}^{R}$, in order to
%build an operator $U_{l}^{R}$.  }}{\Qlb{figure2}}{figure2.eps}%
%{\special{ language "Scientific Word";  type "GRAPHIC";  display "USEDEF";
%valid_file "F";  width 2.1819in;  height 1.4641in;  depth 0pt;
%original-width 0.9599in;  original-height 0.8423in;  cropleft "0";
%filename 'figure2.eps';file-properties "XNPEU";}} }%
%BeginExpansion
%EndExpansion
Finally, to make sure that the operator $U_{l}^{R}$ is Hermitian, we simply
define it as the Hermitian part of $\widehat{U}_{l}^{R}$:%
\begin{equation}
U_{l}^{R}(1,2,...,l)=\frac{1}{2}\left[  \widehat{U}_{l}^{R}(1,2,...,l)+\left(
\widehat{U}_{l}^{R}(1,2,...,l)\right)  ^{\dagger}\right]  \label{a17quat}%
\end{equation}
(we will nevertheless see that, in practice, this symmetrization has no
consequence on the following calculations, which deal only with partial
traces: the distinction between $\widehat{U}_{l}^{R}$ and $U_{l}^{R}$ is not
essential here).

\subsection{Partial trace}

\label{partial}

We now study the following partial trace with respect to particles $2$, $3$,
..., $l$:
\begin{equation}
Tr_{2,3,...l}\left\{  U_{l}^{R}(1,2,3,...,l)\right\}  \label{a18}%
\end{equation}
which is still an operator in the space of states of particle $1$. Since
$U_{l}^{R}$ is obtained from (\ref{a17}) by adding $\sqrt{U_{1}}$'s at
appropriate places in each term of the sum, the trace contained in (\ref{a18})
can itself be expressed as a sum of traces of products of operators.\ To each
of these terms, we will now associate another sort of tree diagram, which
resembles that of figure \ref{figure2}, but where the order of the branches
now characterize the order of operators under the trace (instead of being
related to the numbering of the particles).

\subsubsection{Normal ordering}

\label{normal}

The purpose of this section is to put the operators in a standard order that
can be described by a tree-diagram corresponding to a well-defined
contribution to the value of the partial trace that provides $\rho_{1}$.\ As
before, the open circle at the root of the tree corresponds to particle $1$,
but now the upper branch corresponds to the first sequence of operators that
occur inside the trace, the first sub-branch to the second sequence, etc. This
new tree is actually not very different from the initial tree: the only
difference is actually the way in which the ordering of its branches is
defined.%
%TCIMACRO{\FRAME{ftbpFU}{1.7391in}{1.4589in}{0pt}{\Qcb{This figure shows a
%$\rho_{1}$ tree-diagram where the order of the branches now determines the
%order of operators inside a trace over variables of all particles, except
%particle number $1$; the first vertical branch corresponds to the first group
%of operators, the first secondary branch on the right to the next group,
%etc.\ Moreover, each $\overline{U}_{2}$ carries with it two $\sqrt{U_{1}}$
%operators; in order to simplify the diagrams, we represent them as
%half-squares (rectangles). A full square represents a $U_{1}$ operator.}%
%}{\Qlb{figure3}}{figure3.eps}{\special{ language "Scientific Word";
%type "GRAPHIC";  display "USEDEF";  valid_file "F";  width 1.7391in;
%height 1.4589in;  depth 0pt;  original-width 1.5817in;
%original-height 0.7567in;  cropleft "0";  croptop "1";  cropright "1";
%cropbottom "0";  filename 'figure3.eps';file-properties "XNPEU";}} }%
%BeginExpansion
\begin{figure}
[ptb]
\begin{center}
\includegraphics[
height=1.4589in,
width=1.7391in
]%
{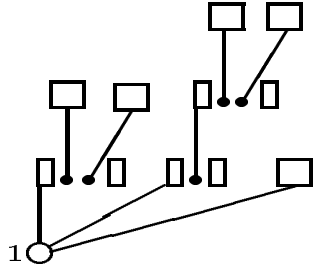}%
\caption{This figure shows a $\rho_{1}$ tree-diagram where the order of the
branches now determines the order of operators inside a trace over variables
of all particles, except particle number $1$; the first vertical branch
corresponds to the first group of operators, the first secondary branch on the
right to the next group, etc.\ Moreover, each $\overline{U}_{2}$ carries with
it two $\sqrt{U_{1}}$ operators; in order to simplify the diagrams, we
represent them as half-squares (rectangles). A full square represents a
$U_{1}$ operator.}%
\label{figure3}%
\end{center}
\end{figure}
%EndExpansion

Let us first consider a given term contained in the right hand side of
(\ref{a17}), and suppose that we are interested in its partial trace - for the
moment we leave aside the $\sqrt{U_{1}}$'s, and define the notion of
\ ``normal ordering'' for the partial trace of any term that appear in the sum
defining $\overline{U}_{l}(1,2,...l)$. To reach this normal ordering, the
first step is to locate the first operator of the product that contains
particle $1$, as well as some other particle $i$, $\overline{U}_{2}(1,i)$, and
to move it to the front (the left side of the product): all operators that
occurred before are moved to the end of the series (the right), in the same
order, by using the property of circular permutation under the trace.\ The
second step is to locate, among all operators now sitting after this operator,
which is the first operator $\overline{U}_{2}(i,j)$ \ containing $i$, and to
move it to the second position, directly to the right of $\overline{U}%
_{2}(1,i)$; this is possible since, if the intermediate operators do not
contain\footnote{They cannot contain particle $i$ by assumption.} particle
$1$, they can be moved to the left of $\overline{U}_{2}(1,i)$ and then to the
end of the series; if they contain particle $1$, they cannot
contain\footnote{They cannot contain both particles $1$ and $j$, since
otherwise the two particles would be linked twice, which is contradictory with
the tree structure of figure \ref{figure2}.} particle $j$, so that they can be
moved just after $\overline{U}_{2}(i,j)$.\ The series now begins with the
product $\overline{U}_{2}(1,i)\overline{U}_{2}(i,j)$.\ The third step is
similar to the second: one locates on the right of $\overline{U}_{2}(i,j)$ the
first operator that contains particle $j$, $\overline{U}_{2}(j,k)$, and moves
it directly to the third rank, by the same method: if the intermediate
operators contain\footnote{They cannot contain particle $j$ by assumption.}
neither particle $1$ nor particle $i$, nor particle $j$, they are moved to the
front and then to the end; if they contain one of them, they do not
contain\footnote{The intermediate operators can not contain at the same time
particle $k$ and either particle $1$, or $i$ or $j$: for instance, if they
contained $k$ and $i$, those two particles would be linked twice (directly and
through particle $j$), which is contradictory with the structure of the
tree-diagram.} particle $k$, they are moved just after $\overline{U}_{2}%
(j,k)$.\ The series now begins with the product $\overline{U}_{2}%
(1,i)\overline{U}_{2}(i,j)\overline{U}_{2}(j,k)$. The same process continues
by iteration until, at some point, one reaches the end of the branch of the
tree, with a numbered particle that does not occur any other $\overline{U}%
_{2}$.

One then proceeds to construct a new branch, and therefore to select a
ramification point.\ For this purpose, one goes backwards from the end along
the first branch, locates the first numbered particle $m$, in operator
$\overline{U}_{2}(m,p)$, that occurs (at least) a second time in the list of
remaining operators, in $\overline{U}_{2}(m,n)$; this latter operator is then
moved directly to the right of $\overline{U}_{2}(m,p)$, and creates the
starting point of another branch. The new branch is then extended by the same
method as the first.\ When this branch is also finished, two cases may occur:
either particle $m$ occurs a third time (or more), so that three branches (or
more) of the tree will originate from the same point; or particle $m$ does not
occur anymore, and one continues to move backwards in the main branch to find
another particle that occurs again in one of the non-ordered $\overline{U}%
_{2}$'s.\ At some point, all operators have been moved to their appropriate
place, and the process stops.

Now, to get the contribution to the trace, we have to add $\sqrt{U_{1}}$'s at
the appropriate places, but this does not change much to the reasoning that me
have made for reaching the normal ordering of the operators: the $\sqrt{U_{1}%
}$'s at the end of the product do not move during the operation, while those
that directly follow $\overline{U}_{2}$'s move together with this operator
(this is possible since they do not change the commutations
rules).\ Consequently, all operators can be put into their normal ordering
exactly in the same way, as for $\overline{U}_{2}$'s only.

Obviously, the numbering of all particles, except particle $1$, is totally
irrelevant for the value of the trace, since it defines dummy variables; what
is important is the geometry of the tree, since it determines the value of the
operator obtained after the partial trace is taken\footnote{But different
trees do not necessarily correspond to different results.}. Therefore, a
diagram such as that of figure \ref{figure3} corresponds to a well-defined
contribution to the trace.\ Another remark is that the contribution of
$\widehat{U}_{l}^{R}$ to the trace already provides an Hermitian operator
acting on particle $1$: by circular permutation under the traces, it is easy
to see that the only change induced by a reversing of the order of the
operators leads to a thee where the order of the branches is reversed, in
other words to a tree that already exists in $\widehat{U}_{l}^{R}$ and ensures
Hermiticity of the partial trace.\ This means that the Hermitian
symmetrization of (\ref{a17quat}) is actually not necessary; from now on, we
will therefore only consider the trace of $\widehat{U}_{l}^{R}$.

\subsubsection{Weights of tree-diagrams}

\label{weights}

{}From the results of the preceding section, we can express the partial trace
of $\widehat{U}_{l}^{R}$ as a sum over all diagrams such as that of figure
\ref{figure3}; we now wish to know what their weight is, or in other words,
how many terms of the double sum (\ref{a17}) correspond to each diagram.\ The
reasoning is analogous to the reasoning of \S\ \ref{classical-tree}, except
that now we are dealing with order of operators, instead of particle
numberings.\ Suppose that we ascribe arbitrary particle numbers to all nodes
of this diagram - since particle $1$ is always at the root of the diagram,
this can be done in $(l-1)!$ different ways.\ To each of these numberings
corresponds a given sequence of $\overline{U}_{2}$ operators, and therefore a
given term of the first summation of (\ref{a17}).\ Now, how many terms of the
second summation then correspond to this given diagram?\ To answer this
question depends on the branching (or ramification) factors $r_{1}$, $r_{2}$,
etc. defined in \S\ \ref{classical-tree}.\ In the list of $\overline{U}_{2}%
$'s, the only non-commuting operators are those which contain one common
particle\footnote{They never contain two, otherwise the diagram would not be
minimally connected.}.\ Starting from any order of operators, and after
ordering them according to the procedure of \S\ \ref{normal}, it is easy to
see that there is a probability $1/(r_{1})!$ that the first ramification will
take place with the appropriate order of chain of operators, a probability
$1/(r_{1})!(r_{2})!$ that the right order is still obtained after two
ramifications, etc.. Finally, among all operators that are contained in the
second sum of (\ref{a17}), the number or terms that correspond to each
tree-diagram is:
\begin{equation}
\frac{(l-1)!}{\Pi_{i}(r_{i})!} \label{a19}%
\end{equation}
If we take into account the factor $(l-1)!$ mentioned before (corresponding to
the first summation) as well as the factor $1/(l-1)!$ which is explicit in
(\ref{a17}), we obtain the following weight of the tree-diagram:
\begin{equation}
(l-1)!\frac{1}{(l-1)!}\frac{(l-1)!}{\Pi_{i}(r_{i})!}=\frac{(l-1)!}{\Pi
_{i}(r_{i})!} \label{a20}%
\end{equation}
so that we eventually obtain:%
\begin{equation}
Tr_{2,3,...l}\left\{  U_{l}^{R}(1,2,3,...,l)\right\}  =\sum_{\left\{  \rho
_{1}\,\,diagrams\right\}  }\frac{(l-1)!}{\Pi_{i}(r_{i})!}\,T_{diag.}(1)
\label{a20bis}%
\end{equation}
where $T_{diag.}(1)$ is the operator (acting on the variables of particle $1$)
that is obtained by the partial trace corresponding to the diagram with
branching factors $r_{1}$, $r_{2}$, etc.

\subsection{Energy shift for $\rho_{I}$}

\label{energy}

The next step is to obtain the one-particle density from the previous
considerations. Its expression in terms of Ursell operators
\cite{Ursell-1-bis} (only retaining their tree-reducible part) is:%

\begin{equation}%
\begin{split}
\rho_{I}(1)  &  =zU_{1}(1)+z\,^{2}Tr_{2}\left\{  U_{2}^{R}(1,2)\,\right\}
+\frac{z^{3}}{2!}\,Tr_{2,3}\left\{  U_{3}^{R}(1,2,3)\,\right\}  +..\\
&  +\frac{z^{l-1}}{(l-1)!}\,Tr_{2,..,l}\left\{  U_{l}^{R}(1,...,l)\,\right\}
+...)+...
\end{split}
\label{a21}%
\end{equation}
Replacing the reduced Ursell operators by their expression as sums of
tree-diagrams and taking the counting factors into account (\ref{a20}), we see
that $\rho_{I}$ becomes:
\begin{equation}%
\begin{split}
\rho_{I}(1)  &  =\sqrt{zU_{1}(1)}\left\{  1+\frac{z}{1!}\times%
\begin{picture}
(10,20) \put(2,2){\circle{4}} \put(-0.5,16){\frame{\makebox(5,5){\ }}%
}\drawline(2,4)(2,16)
\end{picture}
+\frac{z^{2}}{2!}\left(  \frac{2!}{2!}\times%
\begin{picture}
(20,20) \put(2,2){\circle{4}} \put(-0.5,16){\frame{\makebox(5,5){\ }}}
\drawline(2,4)(2,16) \put(3.4,3.4){\line(1,1){12}}\put(13,15){\frame{\makebox
(5,5){\ }}}
\end{picture}
+\frac{2!}{1!}\times%
\begin{picture}
(20,30) \put(2,2){\circle{4}} \put(2,16){\frame{\makebox(2,5){\ }}}
\put(3.5,32){\frame{\makebox(5,5){\ }}} \put(8,16){\frame{\makebox(2,5){\ }}}
\drawline(2,4)(2,16) \drawline(6,17)(6,32) \put(6,17){\circle*{2}}%
\end{picture}
\right)  +...\right. \\
&  \left.  +\frac{z^{l-1}}{(l-1)!}\times\frac{(l-1)!}{\Pi_{i}(r_{i})!}\times%
\begin{picture}
(60,50) \put(2,2){\circle{4}} \put(-2,2){\makebox(0,0){$\scriptscriptstyle1$}}
\put(2,16){\frame{\makebox(2,5){\ }}}\put(4,32){\frame{\makebox(5,5){\ }}}
\put(15,16){\frame{\makebox(2,5){\ }}}\put(14.5,31.5){\frame{\makebox
(5,5){\ }}} \drawline(2,4)(2,16) \drawline(6.5,17)(6.5,32) \put
(6.5,17){\circle*{2}} \put(10,17){\line(1,2){7}}\put(10,17){\circle*{2}}
\put(3.4,3.4){\line(3,2){19}} \put(23,16){\frame{\makebox(2,5){\ }}}
\put(30,16){\frame{\makebox(2,5){\ }}} \put(27.5,17){\line(0,1){15}}
\put(27.5,17){\circle*{2}} \put(27.5,32){\frame{\makebox(2,5){\ }}}
\put(40,32){\frame{\makebox(2,5){\ }}} \put(32,33){\line(0,1){15}}
\put(32,33){\circle*{2}} \put(30,48){\frame{\makebox(5,5){\ }}} \put
(35,33){\line(1,2){7.5}} \put(35,33){\circle*{2}}\put(39.5,48){\frame{\makebox
(5,5){\ }}} \put(3.8,2.2){\line(3,1){40}}\put(41,16){\frame{\makebox(5,5){\ }%
}}
\end{picture}
+...\right\}  \sqrt{zU_{1}(1)}%
\end{split}
\label{a22}%
\end{equation}
It is now possible to regroup diagrams according to the ``first branching
factor'' $1/(r_{1})!$, and to write:
\begin{equation}%
\begin{array}
[c]{ll}%
\rho_{I}(1)= & \sqrt{zU_{1}(1)}\left\{  1+\frac{1}{1!}\left(  z%
\begin{picture}
(20,20) \put(2,2){\circle{4}} \put(-0.5,16){\frame{\makebox(5,5){\ }}%
}\drawline(2,4)(2,16)
\end{picture}
+z^{2}%
\begin{picture}
(20,30) \put(2,2){\circle{4}} \put(2,16){\frame{\makebox(2,5){\ }}}
\put(3.5,32){\frame{\makebox(5,5){\ }}} \put(8,16){\frame{\makebox(2,5){\ }}}
\drawline(2,4)(2,16) \drawline(6,17)(6,32)\put(6,17){\circle*{2}}
\end{picture}
+\frac{1}{2!}z^{3}%
\begin{picture}
(20,30) \put(2,2){\circle{4}} \put(-2,2){\makebox(0,0){$\scriptscriptstyle1$}}
\put(2,16){\frame{\makebox(2,5){\ }}}\put(4,32){\frame{\makebox(5,5){\ }}}
\put(15,16){\frame{\makebox(2,5){\ }}}\put(14.5,31.5){\frame{\makebox
(5,5){\ }}} \drawline(2,4)(2,16) \drawline(6.5,17)(6.5,32) \put
(6.5,17){\circle*{2}} \put(10,17){\line(1,2){7}}\put(10,17){\circle*{2}}
\end{picture}
+...\right)  \right. \\
& \left.  +\frac{1}{2!}\left(  z^{2}%
\begin{picture}
(20,20) \put(2,2){\circle{4}} \put(-0.5,16){\frame{\makebox(5,5){\ }}}
\drawline(2,4)(2,16) \put(3.4,3.4){\line(1,1){12}} \put(13,15){\frame{\makebox
(5,5){\ }}}%
\end{picture}
+...\right)  +...\right\}  \sqrt{zU_{1}(1)}%
\end{array}
\label{a23}%
\end{equation}
Again, we notice a self-similarity property of the expansion, which makes
one-particle density operators appear on the right hand side of the equation:
\begin{equation}
\rho_{I}(1)=\sqrt{zU_{1}(1)}\left\{  1+%
\begin{picture}
(5,20) \put(2,2){\circle{4}}\put(-2,16){\frame{$\scriptscriptstyle{\rho_{I}}$%
}}\drawline(2,4)(2,16)
\end{picture}
+\frac{1}{2!}\times%
\begin{picture}
(25,20) \put(2,2){\circle{4}} \put(-2,16){\frame{$\scriptscriptstyle{\rho_{I}%
}$}}\put(16,0){\frame{$\scriptscriptstyle{\rho_{I}}$}} \drawline(2,4)(2,16)
\drawline(4,2)(16,2)
\end{picture}
+...+\frac{1}{r_{1}!}\times%
\begin{picture}
(25,20) \put(2,2){\circle{4}} \put(-2,16){\frame{$\scriptscriptstyle{\rho_{I}%
}$}} \put(16,0){\frame{$\scriptscriptstyle{\rho_{I}}$}} \put
(14,14){\frame{$\scriptscriptstyle{\rho_{I}}$}}\drawline(2,4)(2,16)
\drawline(4,2)(16,2) \drawline(3,3)(14,14)
\end{picture}
+...\right\}  \sqrt{zU_{1}(1)} \label{a24}%
\end{equation}
We now recognize the development of an exponential operator, and finally
obtain:
\begin{equation}
\rho_{I}(1)=\sqrt{zU_{1}(1)}\exp{(Tr_{2}\left\{  \overline{U}_{2}%
(1,2)\,\rho_{I}(2)\right\}  )}\sqrt{zU_{1}(1)} \label{a25}%
\end{equation}
or:
\begin{equation}
-\beta\Delta(1)=Tr_{2}\left\{  \overline{U}_{2}(1,2)\,\rho_{I}(2)\right\}
\label{a26}%
\end{equation}
We therefore find that the energy shift depends linearly on the one-particle
density operator (proportional to the density), which corresponds exactly to
the mean-field approximation.

\subsection{Physical discussion}

\label{physical}

The preceding calculation illustrates the physics that is behind the
construction of a mean-field: when the test particle $1$ interacts with
another particle $i$, it is also possible that this particle interacts in turn
with particle $j$, and so on.\ Moreover, either the test particle or any other
particle may perfectly well interact with several others; this introduces
branching in the interaction tree, either directly at the root or at any other
place. What is not possible is to create ``loops'': any particle can interact
with one or several neighbors, and these interactions can propagate further to
other particles through many intermediate carriers, but they should never come
back to the original particle.\ This is very similar to the ``no
re-collision'' assumption which is behind the ``molecular chaos Ansatz'' of
the Boltzmann transport equation.

We note that the expression (\ref{a26}) involves the operator $\overline
{U}_{2}(1,2)$, instead of the potential itself $V_{12}$ as in the usual
expressions of the mean-field.\ In other words, the mean-field involves matrix
elements of an exponential containing the interaction potential (see
definition (\ref{a4}) of the second Ursell operator), more precisely the
difference between two exponentials (corresponding so to say to a change in
the local Boltzmann equilibrium), instead of not merely the matrix elements of
$V_{12}$ itself.\ Of course, if this potential can be treated to first order,
this makes no difference, as shown by an elementary calculation.\ But
realistic interatomic potentials can not be treated properly by first order
perturbation theory (Born approximation), and this exponential may actually
introduce an enormous difference.\ A well-known illustration of this fact is
given by alkali atoms, which have a strongly attractive potential sustaining
many bound states ($V_{12}$ is negative, except in the very short range part
of the potential); nevertheless, the phase shift at zero energy may
correspond, either to a positive scattering length (effective repulsion), or
to a negative value (effective attraction), depending on small details of the
potential and on a very delicate balance effect between attraction and
repulsion\footnote{Even a purely attractive potential may have a positive or
negative scattering length, depending on the position of the last bound state
with respect to the continuum.}.\ A naive reasoning in terms of the potential
itself could lead to the idea that, in a dilute gas of alkali atoms, the
mean-field should always be very attractive.\ Of course, this is known to be
incorrect: the mean-field is actually repulsive when $a$ is positive, and
attractive only when $a$ is negative.\ The usual way to understand this
property is to replace the real potential $V_{12}$ by a pseudo-potential,
which is directly proportional to the scattering length $a$ and treated to
first order, a somewhat heuristic method (since the exact reason why using the
real potential is incorrect is not so clear). Here, we clearly see that what
appears naturally is the matrix elements of $\overline{U}_{2}(1,2)$; as shown
in \cite{Ursell-1-ter}, the latter can be expressed in terms of phase shifts
and be shown to be proportional to the scattering length\footnote{More
precisely, this is true for the dominant part of the matrix element, which
depends on the value of the collision wave functions outside of the
interaction potential (asymptotic value of the scattering states).\ Another
contribution arises from the wave functions inside the
potential.\ Nevertheless, if the range of the potential is very small, the
latter contribution remains negligible.} $a$: for instance, if the latter is
positive, we directly get a positive value for the mean-field, without any
special manipulation\footnote{Of course, other methods to prove the same
result also exist; for instance, in the Green's function formalism, the
summation of an infinite series of ladder diagrams can be used to construct
the scattering length from the potential itself.}.

Another remark is that (\ref{a25}) and (\ref{a26}) give $\rho_{I}$ as the
product of exponentials, and not the exponential of a sum, which is not the
same thing if the operators do not commute.\ For translationally invariant
systems, this distinction vanishes since all the single particle operators are
diagonal in the same basis (the momentum basis), and therefore commute.\ In
other cases, one should be careful to take into account non-commutativity of
operators; for instance, the Hermitian operator $\Delta$ can not be seen
exactly as a correction to the one-particle Hamiltonian.

Finally, we remark that our reasoning could be generalized.\ Here, we have
expressed all reducible parts of Ursell operators in terms of $\overline
{U}_{2}$, which is its own irreducible part; we have left aside the
irreducible part $\overline{U}_{3}^{Irr.}$ of $\overline{U}_{3}$.\ But it
should be possible to go further, and to use $\overline{U}_{3}^{Irr.}$ as the
starting point of another decomposition of all $\overline{U}_{l}$'s for
$l\geq4$; in this way, all these $\overline{U}_{l}$'s could give a
contribution to the energy shift $\Delta$ that is quadratic in the
density.\ More details on this calculation are given in Appendix III.

\section{Quantum statistical physics: identical particles}

\label{quantum-stat}

The formalism for identical (Bose or Fermi) particles differs from Boltzmann
particles by the inclusion of exchange cycles.\ They introduce horizontal
parts in the Ursell operator diagrams \cite{Ursell-1, Ursell-1-bis, Ursell-2},
which combine with the vertical lines associated with the Ursell operators;
many more diagrams have to be taken into account in order to include quantum
statistics.\ Despite this big difference, we will see that the formalism
introduces almost the same mechanism and that the exponential of energy
corrections also appear naturally.

\subsection{General equations}

As shown in reference \cite{Ursell-1-bis}, for identical particles, equation
(\ref{a6}) should be replaced by:%
\begin{equation}%
\begin{split}
\rho_{I}=f_{1}  &  +%
\begin{picture}
(60,20) \put(0,0){\frame{$\scriptstyle{1+\eta f_{1}}$}} \put
(35,0){\frame{$\scriptstyle{1+\eta f_{1}}$}}\put(35,15){\frame{$\scriptstyle
{1+\eta f_{1}}$}} \put(23,4){\line(1,0){12}}\put(23,16){\line(0,1){4}}
\put(23,18){\line(1,0){12}} \put(28,4){\line(0,1){14}} \put(30,4){\line
(0,1){14}}
\end{picture}
\\
&  +%
\begin{picture}
(80,20) \put(0,0){\frame{$\scriptstyle{1+\eta f_{1}}$}}\put
(35,0){\frame{$\scriptstyle{1+\eta f_{1}}$}} \put(70,0){\frame{$\scriptstyle
{1+\eta f_{1}}$}} \put(23,4){\line(1,0){12}} \put(58,4){\line(1,0){12}}
\put(28,4){\line(0,1){14}} \put(30,4){\line(0,1){12}}\put(62,4){\line
(0,1){12}} \put(64,4){\line(0,1){14}} \put(28,18){\line(1,0){36}}
\put(30,16){\line(1,0){32}}
\end{picture}
\\
&  +%
\begin{picture}
(90,40) \put(0,0){\frame{$\scriptstyle{1+\eta f_{1}}$}}\put
(35,0){\frame{$\scriptstyle{1+\eta f_{1}}$}} \put(70,15){\frame{$\scriptstyle
{1+\eta f_{1}}$}} \put(35,15){\frame{$\scriptstyle{1+\eta f_{1}}$}}
\put(70,30){\frame{$\scriptstyle{1+\eta f_{1}}$}} \put(23,4){\line(1,0){12}}
\put(58,33){\line(1,0){12}} \put(28,4){\line(0,1){14}}\put(30,4){\line
(0,1){14}} \put(62,18){\line(0,1){15}}\put(64,18){\line(0,1){15}}
\put(23,18){\line(1,0){12}} \put(58,18){\line(1,0){12}}\put(23,16){\line
(0,1){4}} \put(58,31){\line(0,1){4}}
\end{picture}
+...
\end{split}
\label{a27}%
\end{equation}
where $f_{1}$ is the one-particle density operator for an ideal gas:%
\begin{equation}
f_{1}(1)=\frac{zU_{1}(1)}{1-\eta\,zU_{1}(1)} \label{b2}%
\end{equation}
with $\eta=+1$ for bosons, $\eta=-1$ for fermions.\ In (\ref{a27}), the
operators $\left[  1+\eta f_{1}\right]  $ arise from a summation over all
possible values of cycle length $l$ (ranging from $l=0$ to infinity) of the
product $\left[  \eta zU_{1}(1)\right]  ^{l}$ - the origin of the factor
$\eta$ is that, for fermions, every exchange cycle of length $l$ introduces a
$(-1)^{l}$ sign into the diagram which contains it. Generally speaking, in all
such diagrams, the lowest horizontal line corresponds to the multiplication of
operators acting in the space of particle $1$, while all the other lines above
correspond to traced variables of other particles.

In \cite{Ursell-2}, the series giving $\rho_{I}$ was re-written in a
self-consistent way, which resums in one single diagram an infinite number of
diagrams of the initial series (\ref{a27}):%

\begin{equation}%
\begin{split}
\rho_{I}=f_{1}  &  +%
\begin{picture}
(60,20) \put(0,0){\frame{$\scriptstyle{1+\eta f_{1}}$}} \put
(35,0){\frame{$\scriptstyle{1+\eta\rho_{I}}$}}\put(35,15){\frame{$\scriptstyle
{1+\eta\rho_{I}}$}} \put(23,4){\line(1,0){12}}\put(23,16){\line(0,1){4}}
\put(23,18){\line(1,0){12}} \put(28,4){\line(0,1){14}} \put(30,4){\line
(0,1){14}}
\end{picture}
\\
&  +%
\begin{picture}
(80,20) \put(0,0){\frame{$\scriptstyle{1+\eta f_{1}}$}}\put
(35,0){\frame{$\scriptstyle{1+\eta\rho_{I}}$}} \put(70,0){\frame{$\scriptstyle
{1+\eta\rho_{I}}$}} \put(23,4){\line(1,0){12}} \put(58,4){\line(1,0){12}}
\put(28,4){\line(0,1){14}} \put(30,4){\line(0,1){12}}\put(62,4){\line
(0,1){12}} \put(64,4){\line(0,1){14}} \put(28,18){\line(1,0){36}}
\put(30,16){\line(1,0){32}}
\end{picture}
\\
&  +%
\begin{picture}
(100,25) \put(0,0){\frame{$\scriptstyle{1+\eta f_{1}}$}}\put
(35,0){\frame{$\scriptstyle{1+\eta\rho_{I}}$}} \put(70,0){\frame{$\scriptstyle
{1+\eta\rho_{I}}$}} \put(35,15){\frame{$\scriptstyle{1+\eta\rho_{I}}$}}%
\put(23,4){\line(1,0){12}} \put(58,4){\line(1,0){12}}\put(28,4){\line
(0,1){14}} \put(30,4){\line(0,1){14}} \put(62,4){\line(0,1){14}}%
\put(64,4){\line(0,1){14}} \put(23,18){\line(1,0){12}}\put(58,18){\line
(1,0){12}} \put(23,16){\line(0,1){4}} \put(70,16){\line(0,1){4}}%
\end{picture}
+...
\end{split}
\label{b0}%
\end{equation}
For instance, the diagram of the third line of (\ref{a27}), as well as many
other similar diagrams, are now included in the first term after $f_{1}$ in
(\ref{b0}), which symbolizes the partial trace:%
\begin{equation}
2z^{2}\left(  1+\eta f_{1}(1)\right)  \,Tr_{2}\left\{  U_{2}(1,2)\left[
1+\eta\rho_{I}(2)\right]  \right\}  \,\left(  1+\eta\rho_{I}(1)\right)
\label{b3}%
\end{equation}
\ In (\ref{b0}), the second term after $f_{1}$ turns out to be the exchange
term of (\ref{b3}), since it can be written:%
\begin{equation}
2z^{2}\left(  1+\eta f_{1}(1)\right)  \,Tr_{2}\left\{  U_{2}(1,2)\,\eta
\,P_{ex.}(1,2)\,\left[  1+\eta\rho_{I}(2)\right]  \right\}  \,\left(
1+\eta\rho_{I}(1)\right)  \label{b4}%
\end{equation}
where $P_{ex.}$ is the exchange operator between particles $1$ and $2$.\ These
two terms therefore group naturally together with the introduction of the
symmetrized ($S$: $\eta=1$) or antisymmetrized ($A$: $\eta=-1$) form of the
$U_{2}$ operator:%
\begin{equation}
U_{2}^{S,A}(1,2)\,=\,\frac{1+\eta P_{ex.}(1,2)}{2}\,\,U_{2}(1,2) \label{b5}%
\end{equation}
\ For a \emph{dilute} gas, they were interpreted in \cite{Ursell-2} as the
mean-field correction, which provides no correction to the critical
temperature of the gas within the MIME approximation (see \S\ref{notation}%
).\ The explicit expression of the third term is, similarly:%
\begin{equation}%
\begin{array}
[c]{r}%
2z^{2}\left(  1+\eta f_{1}(1)\right)  \,Tr_{2}\left\{  U_{2}(1,2)\,\left(
1+\eta\rho_{I}(1)\right)  \,\left(  1+\eta\rho_{I}(2)\right)  \times\right. \\
\left.  \times U_{2}(1,2)\,\left(  1+\eta\rho_{I}(2)\right)  \,\left(
1+\eta\rho_{I}(1)\right)  \right\}
\end{array}
\label{b6}%
\end{equation}
and, as the preceding term, it groups with another diagram (not shown) that
appears as its exchange diagram.%

%TCIMACRO{\FRAME{ftbpFU}{1.5056in}{1.4477in}{0pt}{\Qcb{With quantum statistics,
%this diagram replaces the diagram in the right part of figure \ref{figure2};
%note the presence of the statistical $[1+\eta\rho_{I}]$ factors.}%
%}{\Qlb{figure4}}{figure4.eps}{\special{ language "Scientific Word";
%type "GRAPHIC";  display "USEDEF";  valid_file "F";  width 1.5056in;
%height 1.4477in;  depth 0pt;  original-width 0.8475in;
%original-height 0.5673in;  cropleft "0";  croptop "1";  cropright "1";
%cropbottom "0";  filename 'figure4.eps';file-properties "XNPEU";}} }%
%BeginExpansion
\begin{figure}
[ptb]
\begin{center}
\includegraphics[
height=1.4477in,
width=1.5056in
]%
{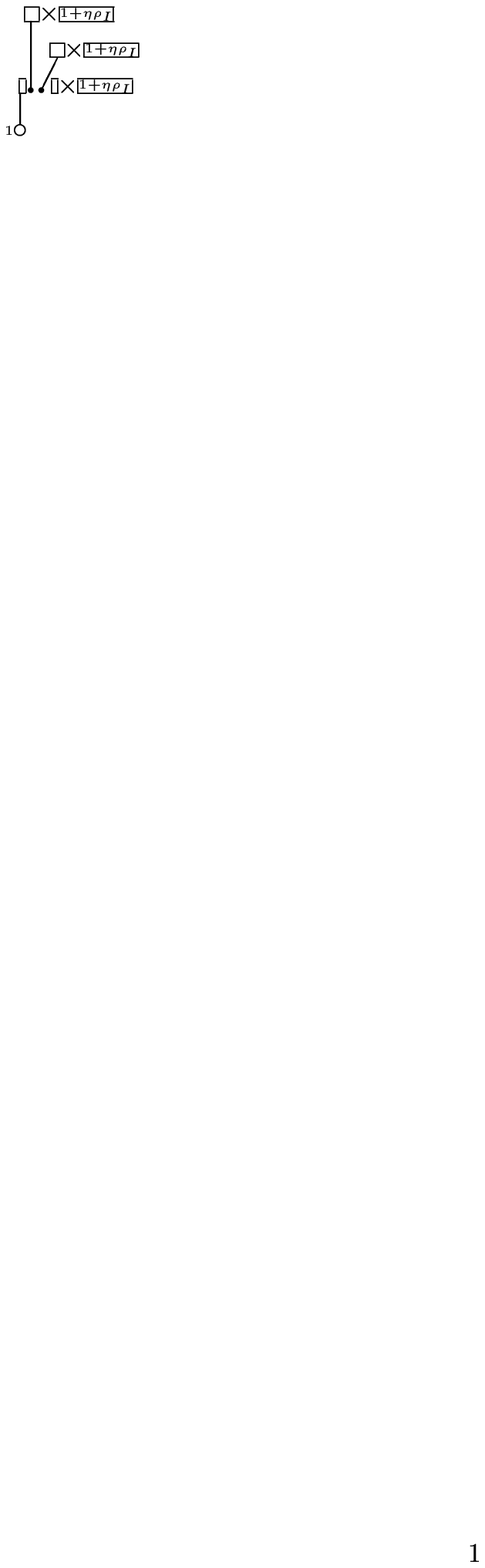}%
\caption{With quantum statistics, this diagram replaces the diagram in the
right part of figure \ref{figure2}; note the presence of the statistical
$[1+\eta\rho_{I}]$ factors.}%
\label{figure4}%
\end{center}
\end{figure}
%EndExpansion

\subsection{Introducing exponentials}

Our purpose now is to introduce exponentials in order to make energy shifts
appear explicitly.\ For the ideal gas, the one particle distribution is given
by (\ref{b2}) so that, since $\eta^{2}=1$:%
\begin{equation}
1+\eta\,(f_{1})^{-1}=\eta\,\,\left[  zU_{1}\right]  ^{-1}=\eta\exp\left[
\beta\,(H_{1}(1)-\mu)\right]  \label{b6ter}%
\end{equation}
Similarly:%
\begin{equation}
\eta\,\left[  1+\eta\,(f_{1})^{-1}\right]  ^{-1}=\frac{f_{1}}{1+\eta f_{1}%
}=z\,U_{1} \label{b6quater}%
\end{equation}
where the same exponential appears, now with a negative exponent. To calculate
energy shifts, it is therefore natural to introduce quantities such as
$1+\eta(\rho_{1})^{-1}$ or $\rho_{1}/(1+\eta\rho_{1})$.

To do this, we remark that the lower lines of all diagrams contained in
(\ref{a27}) start and end with\ the same operators, so that equation
(\ref{b0}) can be re-written as:%
\begin{equation}
\rho_{I}(1)=f_{1}(1)+\left[  1+\eta f_{1}(1)\right]  \,\,K(1)\,\,\left[
1+\eta\rho_{I}(1)\right]  \label{b7}%
\end{equation}
where the operator $K$ is an infinite sum of partial traces, containing
various number of traced $U_{2}$'s as well as $\left[  1+\eta\rho
_{I}(i)\right]  $'s in the horizontal lines:%
\begin{equation}%
\begin{split}
K=%
\begin{picture}
(30,20) \put(12,13){\frame{$\scriptstyle{1+\eta\rho_{I}}$}}\put(0,2){\line
(1,0){12}} \put(0,14){\line(0,1){4}} \put(0,0){\line(0,1){4}}\put
(12,0){\line(0,1){4}} \put(0,16){\line(1,0){12}} \put(5,2){\line(0,1){14}}%
\put(7,2){\line(0,1){14}}
\end{picture}
&  +%
\begin{picture}
(50,20) \put(12,-1){\frame{$\scriptstyle{1+\eta\rho_{I}}$}} \put
(0,2){\line(1,0){12}}\put(35,2){\line(1,0){12}} \put(5,2){\line(0,1){14}}%
\put(7,2){\line(0,1){12}}\put(39,2){\line(0,1){12}} \put(41,2){\line(0,1){14}%
}\put(5,16){\line(1,0){36}} \put(7,14){\line(1,0){32}} \put(0,0){\line
(0,1){4}}\put(47,0){\line(0,1){4}}
\end{picture}
\\
&  +%
\begin{picture}
(50,30) \put(12,13){\frame{$\scriptstyle{1+\eta\rho_{I}}$}}\put(0,2){\line
(1,0){12}} \put(0,14){\line(0,1){4}} \put(0,0){\line(0,1){4}}\put
(12,0){\line(0,1){4}} \put(0,16){\line(1,0){12}} \put(5,2){\line(0,1){14}}%
\put(7,2){\line(0,1){14}} \put(35,16){\line(1,0){12}}\put(35,30){\line
(1,0){12}} \put(40,16){\line(0,1){14}} \put(42,16){\line(0,1){14}}
\put(35,28){\line(0,1){4}} \put(47,13){\frame{$\scriptstyle{1+\eta\rho_{I}}$}}
\put(47,27){\frame{$\scriptstyle{1+\eta\rho_{I}}$}}%
\end{picture}
\\
&  +%
\begin{picture}
(50,20) \put(12,-1){\frame{$\scriptstyle{1+\eta\rho_{I}}$}}\put
(12,14){\frame{$\scriptstyle{1+\eta\rho_{I}}$}} \put(0,2){\line(1,0){12}}%
\put(35,2){\line(1,0){12}} \put(5,2){\line(0,1){15}} \put(7,2){\line(0,1){15}%
}\put(39,2){\line(0,1){15}} \put(41,2){\line(0,1){15}}\put(0,17){\line
(1,0){12}} \put(35,17){\line(1,0){12}} \put(0,15){\line(0,1){4}}%
\put(47,15){\line(0,1){4}} \put(0,0){\line(0,1){4}} \put(47,0){\line(0,1){4}}%
\end{picture}
+...
\end{split}
\label{b8}%
\end{equation}
In other words, the diagrams corresponding to $K$ are obtained from those
corresponding to $\rho_{I}$ by simply removing the two external parts of the
lowest line (diagram amputation). The operator $K$ is clearly Hermitian, as
can be shown by using circular permutation of operators under the
trace.\ There is an obvious analogy between operator $K$ and the self-energies
in the formalism of Green's functions.

Now, if we multiply both sides of (\ref{b7}) by $\left(  f_{1}\right)  ^{-1}$
on the left, and by $(\rho_{I})^{-1}$ on the right, we obtain:%
\begin{equation}
\left(  f_{1}\right)  ^{-1}=(\rho_{I})^{-1}+\left(  1+\eta\,\left(
f_{1}\right)  ^{-1}\right)  \,K\,\left(  1+\eta\,\left(  \rho_{I}\right)
^{-1}\right)  \label{b8bis}%
\end{equation}
or, if we multiply both sides by $\eta$, add $1$, and take (\ref{b6quater})
into account:%
\begin{equation}
1+\eta\,\left(  f_{1}\right)  ^{-1}=\left[  1+\left(  zU_{1}\right)
^{-1}K\right]  \left[  1+\eta\,(\rho_{I})^{-1}\right]  \label{b9}%
\end{equation}
or again:%
\begin{equation}
\frac{\,\rho_{I}}{1+\eta\rho_{I}}=z\,U_{1}+K \label{b9bis}%
\end{equation}
In the absence of interaction, $K$ vanishes and we recover (\ref{b6quater});
$K$ appears therefore as analogous to a correction to $zU_{1}$ introduced by
the interactions, in other words to a correction to the single particle energy.

To make this analogy more precise, by similarity with (\ref{b6ter}), we define
the energy shift operator $\Delta(1)$ by\footnote{We could also have used
another definition of the energy shift operator, by introducing the
exponential of a sum instead of the product of exponentials:
\begin{equation}
\left[  1+(\eta\rho_{I})^{-1}\right]  =\eta\,\,e^{\beta(H_{1}(1)+\Delta
(1)-\mu)}\nonumber
\end{equation}
It turns out that, for the present calculation, definition (\ref{b12}) is more
convenient.}:%
\begin{equation}
1+(\eta\rho_{I})^{-1}=\eta\,\left[  zU_{1}(1)\right]  ^{-1/2}\,e^{\beta
\Delta(1)}\left[  zU_{1}(1)\right]  ^{-1/2} \label{b12}%
\end{equation}
which immediately leads to:%
\begin{equation}
\exp\left\{  -\beta\,\Delta(1)\right\}  =\,\left[  1+\frac{1}{\sqrt{zU_{1}%
(1)}}\,\,K\,(1)\,\frac{1}{\sqrt{zU_{1}(1)}}\right]  \label{b12prime}%
\end{equation}
The series defining the operator $K(1)$ therefore contains all the information
on the energy shifts for the single-particle density operator; equation
(\ref{b12prime}) is the equivalent of relation (\ref{a13}) of Appendix II, in
an operator form, when quantum statistics is added.

\subsection{Calculating the energy shifts}

Our purpose now is to express the energy shift as a function of the density,
or of the single particle density operator $\rho_{I}$, as we did in classical
statistical physics.\ This raises two problems: first, we have to build an
exponential from the infinite series that provides $K$; second, we have to
understand how the density operator $\rho_{I}$ appears, instead of the factors
$\left[  1+\eta\rho_{I}(i)\right]  $ that seem to be systematically present in
all terms of the series.\ We will see that the two problems ``cure each
other'' and that the result the calculation leads to an systematic and
satisfactory grouping of the terms for a dense system. To do this, we begin
with a simple case, where only the direct term of the mean-field is taken into account.

\subsubsection{Direct term}

We now take an approximation of expression (\ref{b8}) of $K$ by retaining only
the diagrams that contain one single $U_{2}$, one single $U_{3}$, ... one
single $U_{l}$, ... and where all the $U_{l}$ operators are connected to $l$
different cycles:%
\begin{equation}
K=%
\begin{picture}
(30,20) \put(2,0){\line(0,1){4}} \put(14,0){\line(0,1){4}}\put(2,2){\line
(1,0){12}} \put(14,11){\frame{$\scriptstyle{1+\eta\rho_{I}}$}}\put
(2,12){\line(0,1){4}} \put(2,14){\line(1,0){12}} \put(7,2){\line(0,1){12}}%
\put(9,2){\line(0,1){12}}
\end{picture}
+%
\begin{picture}
(30,25) \put(2,0){\line(0,1){4}} \put(14,0){\line(0,1){4}} \put(2,2){\line
(1,0){12}}\put(2,12){\line(0,1){4}} \put(14,11){\frame{$\scriptstyle
{1+\eta\rho_{I}}$}}\put(2,14){\line(1,0){12}} \put(2,24){\line(0,1){4}}%
\put(14,23){\frame{$\scriptstyle{1+\eta\rho_{I}}$}} \put(2,26){\line(1,0){12}%
}\put(6,2){\line(0,1){24}} \put(8,2){\line(0,1){24}} \put(10,2){\line
(0,1){24}}%
\end{picture}
+... \label{K-mean}%
\end{equation}
Moreover, we replace each Ursell operators $U_{l}$ by its tree-reducible
part.\ This leads to the equation:
\begin{equation}%
\begin{array}
[c]{ll}%
K(1)= & z^{2}\ Tr_{2}\left\{  U_{2}(1,2)\,\left[  1+\eta\rho_{I}(2)\right]
\right\} \\
& +\frac{z^{3}}{2!}\,Tr_{2,3}\left\{  U_{3}^{R}(1,2,3)\,(\left[  1+\eta
\rho_{I}(2)\right]  \,\left[  1+\eta\rho_{I}(3)\right]  \right\} \\
& +....\\
& +\frac{z^{l}}{(l-1)!}\,Tr_{2,..,l}\left\{  U_{l}^{R}(1,..,l)\,\left[
1+\eta\rho_{I}(2)\right]  ...\times\left[  1+\eta\rho_{I}(l)\right]  \right\}
\\
& +...
\end{array}
\label{b14}%
\end{equation}
For the sake of simplicity, we will even replace here the operators $U_{l}%
^{R}$ by their non-symmetrized part $\widehat{U}_{l}^{R}$ (see equation
(\ref{a17quat})); nevertheless, since at the end we will find an Hermitian
operator, this has no consequence on the result.\ Graphically, if we represent
as before the $\sqrt{U_{1}}$'s by vertical rectangles, $U_{1}$'s by squares
and multiplication by $\left[  1+\eta\rho_{1}\right]  $ by long horizontal
rectangles as in figure \ref{figure4}, (\ref{b14}) then corresponds to:
\begin{equation}
K(1)=\frac{z}{1!}\times%
\begin{picture}
(20,20) \put(2,2){\circle{4}}\put(-0.5,16){\frame{\makebox(5,5){\ }}$\times
$\frame{$\scriptscriptstyle{1+\eta\rho_{1}}$}} \drawline(2,4)(2,16)
\end{picture}
+\frac{z^{2}}{2!}(\frac{2!}{2!}\times%
\begin{picture}
(45,20) \put(2,2){\circle{4}}\put(-0.5,16){\frame{\makebox(5,5){\ }}$\times
$\frame{$\scriptscriptstyle{1+\eta\rho_{1}}$}} \drawline(2,4)(2,16)
\put(3.4,3.4){\line(2,1){12}}\put(15.5,7){\frame{\makebox(5,5){\ }}$\times
$\frame{$\scriptscriptstyle{1+\eta\rho_{1}}$}}
\end{picture}
+\frac{2!}{1!}\times%
\begin{picture}
(30,30) \put(2,2){\circle{4}} \put(2,16){\frame{\makebox(2,5){\ }}}%
\put(3.5,32){\frame{\makebox(5,5){\ }}$\times$\frame{$\scriptscriptstyle
{1+\eta\rho_{1}}$}} \put(8,16){\frame{\makebox(2,5){\ }}$\times$%
\frame{$\scriptscriptstyle{1+\eta\rho_{1}}$}} \drawline(2,4)(2,16)
\drawline(6,17)(6,32) \put(6,17){\circle*{2}}%
\end{picture}
)+... \label{b15}%
\end{equation}
We now notice the same kind of self-similarity of the series than in the case
of Boltzmann statistics.\ There are nevertheless two differences.\ The first
is that the operator that is resummed in the branches is now $K$, or actually
the sum $zU_{1}+K$ (the term $zU_{1}$ is introduced by the $1$ that is present
at each node but absent as the first term in the series); the second is that
this operator is multiplied on the right by the factor $\left[  1+\eta
\,\rho_{I}\right]  $, so that we get the product:%
\begin{equation}
\left[  zU_{1}+K\right]  \times\,\left[  1+\eta\,\rho_{I}\right]  \label{b16}%
\end{equation}
But, according to (\ref{b9bis}), this product is merely equal to $\rho_{I}$.
We therefore have:%
\begin{equation}
K(1)=\sqrt{U_{1}(1)}\left[  \exp\,\left(  Tr_{2}\left\{  \overline{U}%
_{2}(1,2)\,\rho_{I}(2)\right\}  \right)  -1\right]  \sqrt{U_{1}(1)}
\label{b17}%
\end{equation}
(the $-1$ arises because the first term of the exponential series is absent
from $K$).\ Inserting this result into (\ref{b12prime}) finally provides:%
\begin{equation}
-\beta\Delta(1)=Tr_{2}\left\{  \overline{U}_{2}(1,2)\rho_{I}(2)\right\}
\label{b19}%
\end{equation}
which is exactly the same result as for Boltzmann statistics. We can now check
that the final result is Hermitian, so that it was indeed correct to ignore
the Hermitian symmetrization of the operators $\widehat{U}_{l}^{R}$.

\subsubsection{Exchange term}

We now add exchange terms into (\ref{K-mean}).\ An example of such a term is
the second term of the right hand side of (\ref{b8}): we have already noticed
that this term is obtained by replacing $\overline{U}_{2}(1,2)$ by
$\overline{U}_{2}(1,2)P_{ex}$ in the direct mean-field term associated with
the linear density term in $K$.\ In other words, the sum of the two terms
corresponds to the following substitution:%
\begin{equation}
\overline{U}_{2}(1,2)\,\Rightarrow\,\overline{U}_{2}(1,2)\,\left[  1+\eta
P_{ex.}(1,2)\right]  =2\,\overline{U}_{2}^{S,A}(1,2) \label{e1}%
\end{equation}
where $P_{ex.}(1,2)$ is the exchange operator for particles $1$ and $2$. More
generally, the question is whether each term in the whole series of terms of
(\ref{b15}), which eventually leads to an exponential, contains all exchange
terms that are necessary to perform substitution (\ref{e1}).%
%TCIMACRO{\FRAME{ftbpFU}{4.7781in}{1.0369in}{0pt}{\Qcb{Exchange diagrams for
%the case $l=3$; the direct term from which these diagrams are obtained is the
%second term in the right hand side of (\ref{K-mean}).}}{\Qlb{figure5}%
%}{figure5.eps}{\special{ language "Scientific Word";  type "GRAPHIC";
%display "USEDEF";  valid_file "F";  width 4.7781in;  height 1.0369in;
%depth 0pt;  original-width 0.8043in;  original-height 0.4125in;
%cropleft "0";  croptop "1";  cropright "1";  cropbottom "0";
%filename 'figure5.eps';file-properties "XNPEU";}} }%
%BeginExpansion
\begin{figure}
[ptb]
\begin{center}
\includegraphics[
height=1.0369in,
width=4.7781in
]%
{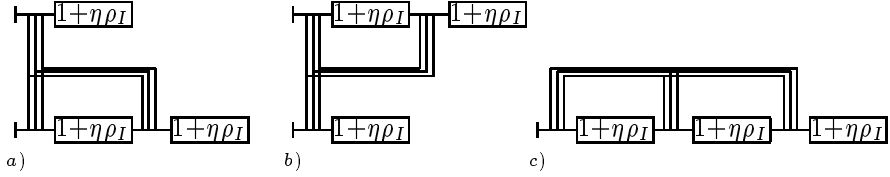}%
\caption{Exchange diagrams for the case $l=3$; the direct term from which
these diagrams are obtained is the second term in the right hand side of
(\ref{K-mean}).}%
\label{figure5}%
\end{center}
\end{figure}
%EndExpansion

In order to construct the diagrams which are exchange diagrams for the
mean-field, it is useful to remember the origin of each Ursell diagram: it
arises from the association of a set of $U_{l}$'s (with $l\geq2$) with
permutation cycles; in the diagrams that we have retained so far, only one
$U_{l}$ is present, and each of the $l$ particles is included in a different
cycle, containing none of the other particles.\ Let us now consider two
particles, $i$ and $k$, that sit initially in different cycles; if we apply an
additional exchange operator $P_{ex}(i,j)$ to the product of the two cycles,
it is easy to see that we obtain a larger cycle with an length which is the
sum of the two initial lengths: the two cycles merely coalesce into one.\ This
immediately leads to another Ursell diagram, such as the two first that are
shown in figure \ref{figure5} (a and b) in the particular case $l=3$.\ One can
then repeat the operation and apply another exchange operator to the result,
which will make two more cycles fuse together into an even larger cycle; in
this way still another diagram is obtained, such as the last shown in figure
\ref{figure5} (c).\ This introduction of exchange may be repeated until all
pairs of particles contained in $\overline{U}_{2}$ links are exchanged, which
leads to a maximum cycle of length $l$ containing all of them.$\;$Altogether,
$2^{l-1}-1$ different diagrams correspond to all possible ways to fuse
together the various cycles; they provide all the exchange diagrams associated
with the initial direct diagram.

Finally, we have to replace in all these diagrams $U_{l}$ by the sum of
product of $\overline{U}_{2}$ operators that corresponds to $U_{l}^{R}$, as we
did for the direct terms; we replace the ``skeleton'' provided by $U_{l}$ by
all possible trees.\ The result of this operation is the same as for the
direct terms: any branching at the root introduce a product of operators,
while branching at the other nodes introduces a product inside a partial
trace.\ The situation is thus not different than before, except that two sorts
of links now occur, with or without exchange; the summation of the series then
provides nothing but the exponential of a sum. We therefore obtain the simple
result: \
\begin{equation}
-\beta\Delta(1)=Tr_{2}\left\{  \overline{U}_{2}(1,2)\,\left[  1+\eta
P_{ex.}(1,2)\right]  \,\rho_{I}(2)\right\}  \label{c1}%
\end{equation}
which shows that it is the symmetrized form of $\overline{U}_{2}$ that appears
naturally in the expression of the mean-field.

\subsubsection{Correlations}

\begin{figure}
[ptb]
\begin{center}
\includegraphics[
height=0.6702in,
width=1.2021in
]%
{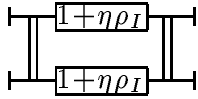}%
\caption{An example of a ``correlation diagram'', where exchange cycles and
Ursell operators ($U_{2}$'s e.g.) combine to make a ``loop''.\ This diagram
provides a contribution which is beyond mean-field.}%
\label{figure6}%
\end{center}
\end{figure}

Can we go further in this exponentiation, and try to include in the energy
shift terms such as that corresponding to the diagram of figure \ref{figure6}?%

%TCIMACRO{\FRAME{ftbpFU}{1.2021in}{0.6702in}{0pt}{\Qcb{An example of a
%``correlation diagram'', where exchange cycles and Ursell operators ($U_{2}$'s
%e.g.) combine to make a ``loop''.\ This diagram provides a contribution which
%is beyond mean-field.}}{\Qlb{figure6}}{figure6.eps}%
%width 1.2021in;  height 0.6702in;  depth 0pt;  original-width 0.7619in;
%original-height 0.4125in;  cropleft "0";  croptop "1";  cropright "1";
%cropbottom "0";  filename 'figure6.eps';file-properties "XNPEU";}} }%
%BeginExpansion
%EndExpansion
As we have seen, the exponentiation operation involves the consideration of
operators that are the square, the cube, etc. of the lowest order operator.
Could we somehow consider diagrams such as that of figure \ref{figure6}, but
where $U_{3}$ are introduced, replace $U_{3}$ by a tree diagram, and show that
the result is the square of the initial operator?\ Actually, this does not
seem to be possible.\ One reason is that the square of an operator such as
that shown in figure \ref{figure6} would involve particle $1$ to be part of
two independent cycles, a situation that never occurs in Ursell diagrams. For
this reason, we have not been able to exponentiate the diagram of figure
\ref{figure6}.\ One possibility is to consider this class of terms as smalls
correction to the mean-field exponential terms, to be treated linearly (as was
done for instance in \cite{Ursell-2}). This is indeed possible above the
Bose-Einstein transition point, but it is certainly not correct below the this
point, as we discuss below.\ As a consequence, the validity of our
calculations is limited to non-condensed Bose systems.

\subsection{Physical discussion}

The energy shift operator $\Delta(1)$ is reminiscent of the self-energy
operator of temperature Green's function's theories, but actually it
corresponds to a different notion. Self-energies include a notion of time, or
frequency (or Matsubara discrete frequencies), while time evolution is absent
from the formalism of Ursell operators: only equilibrium properties are
obtained, which excludes notions such as time dependent response functions,
collective modes and quasiparticles. In other words, only the equilibrium
single particle density operator is calculated, which contains less
information (integrated information over all frequencies) than full Green's
functions.\ It is nevertheless interesting to see how the formalism manages to
reconstruct the energy exponentials, with or without exchange cycles, and that
at the end the mean-field expression simplifies to introduce the single
particle density operator itself in a consistent way.\ It is also worth
noticing that the relevant matrix elements to calculate this mean field are
not those of the bare interaction potential $V_{12}$, but rather those of an
operator $\overline{U}_{2}(1,2)$ that contains $V_{12}$ in an exponential - we
have already discussed in \S\ \ref{physical}, its relation with the
introduction of the scattering length as the relevant parameter to describe
the interactions.

Of course, one should keep in mind that the notion of mean-field is not exact:
we have summed only limited classes of diagrams; moreover, we have replaced
the $U_{l}$'s by their reducible part $U_{l}^{R}$. In the case of bosons for
instance, it is known \cite{Ursell-2, BBHLV} that correlation diagrams such as
that shown in figure \ref{figure6} play an essential role just above the
Bose-Einstein transition point, and even that the theory takes a
non-perturbative character at the transition temperature, so that no
limitation to any finite set of diagrams (or of class of diagrams) is in
principle possible.\ Below the condensation point, the situation is even more
dramatic.\ The reason is that a ladder diagram with $M$ ladders ($U_{2}$
operators) and an intermediate state with an extensive population (the
condensed state) is proportional to the $\mathcal{V}^{M-2}$ (where
$\mathcal{V}$ is the volume), which makes it diverge in the thermodynamic
limit as soon as $M\geq3$: however small the interaction parameter is, these
terms will always dominate the others in the thermodynamic limit. A similar
situation occurs for the so called ``bubble diagrams''.\ As a consequence,
since we have ignored all these correlation diagrams, in their present form
our calculations have a validity that is limited to non-condensed boson
systems.\ Actually, one would expect that, in a condensed system, the
mean-field should include two different parts, due to excitations and to the
condensate respectively, but for the moment we have not explored this question.

\section{Conclusion}

At the end of their 1938 article, Kahn and Uhlenbeck \cite{Kahn} remark that
``[they] have not been able to generalize this physical interpretation of the
$\beta_{l}$ [(irreducible cluster integrals)] to the quantum theory.''\ In the
present article, we have achieved this goal: we have provided a consistent
derivation of the exponentials that introduce the energy shifts $\Delta$, from
which the equation of state can in turn be derived as in classical statistical
physics.\ Actually, in the article, we emphasize how the linear density term
can be derived, but Appendix III discusses briefly how higher order density
terms could also be included.\ For a classical system, our method is
different, and in a sense simpler, than that of Hansen and McDonald
\cite{Hansen}; it does not require any reasoning in the complex plane and we
have shown how the exponentials of $\rho_{I}$ appear rather naturally in the
calculations. The operator that plays the basic role in all our calculations
is $\overline{U}_{2}(1,2)$; its matrix elements depend on the asymptotic (long
distance) properties of the two body scattering wave functions.\ But, in fact,
$\overline{U}_{2}(1,2)$ contains more information than only asymptotic wave
functions and phase shifts: it also contains information about short range
effects of the potential (atoms in the middle of a collision) as well as about
bound states; it would be interesting to explore their consequences on the
properties of the mean-field.

\bigskip\bigskip The Laboratoire Kastler Brossel (LKB) is Unit\'{e} Mixte de
Recherche du CNRS (UMR 8552) et de l'Universit\'{e} Pierre et Marie Curie (Paris).

\bigskip\bigskip

\begin{center}
APPENDIX\ I
\end{center}

In this appendix, we explain why classical tree-diagrams consisting of $l$
particles all have the same numerical value, namely $(\beta_{1})^{l-1}$.Let us
consider the following integral corresponding to a $l$ particles
tree-diagram:
\begin{equation}
T_{diag.}=\frac{1}{\mathcal{V}}\int d^{3}r_{1}\,d^{3}r_{2}...d^{3}%
r_{l}\ f_{12}...f_{ij}...f_{kl}\label{A1}%
\end{equation}
Each Mayer function appearing in it depends on the relative distance between
two particles:%
\begin{equation}
f_{ij}=f(\left|  \mathbf{r}_{i}-\mathbf{r}_{j}\right|  )=f(r_{ij}%
)\label{A1bis}%
\end{equation}
Let $\mathbf{x}_{k}$ be the relative position vector between particle $k$ and
the preceding particle in the branch starting from the root. There is $l-1$
such vectors ($k=2$ to $l$). The important thing to notice is that each Mayer
function depends on only one $\mathbf{x}_{k}$, and that two Mayer functions in
the product necessarily depend on different $\mathbf{x}_{k}$. Let
\begin{equation}
\mathbf{R}=\frac{1}{l}\left(  \mathbf{r}_{1}+\mathbf{r}_{2}+...+\mathbf{r}%
_{l}\right)  \label{A2}%
\end{equation}
be the center of ``mass'' position vector. Then making a change of variables
in the integral from ($\mathbf{r}_{1}$, $\mathbf{r}_{2}$,..., $\mathbf{r}_{l}%
$) to ($\mathbf{R}$, $\mathbf{x}_{2}$,..., $\mathbf{x}_{l}$) (the Jacobian is
one), we obtain:
\begin{equation}
T_{diag.}=\frac{1}{\mathcal{V}}\int d^{3}R\ d^{3}x_{2}...d^{3}x_{l}%
\ f(x_{2})\ ...f(x_{l})\label{A3}%
\end{equation}
We can then integrate over the variable $\mathbf{R}$ to get rid of the volume
factor. The $l-1$ integrals that are left separate so that:%
\begin{equation}
T_{diag.}=\int d^{3}x_{2}\ f(x_{2})...\int d^{3}x_{l}\ f(x_{l})=\left(  \int
d^{3}x\ f(x)\right)  ^{l-1}\label{A4}%
\end{equation}
But the integral of the Mayer function is precisely what we called the first
irreducible cluster integral $\beta_{1}$ (see equation (\ref{12bis})), so that
we finally get:%
\begin{equation}
T_{diag.}=(\beta_{1})^{l-1}\label{A5}%
\end{equation}

\bigskip\bigskip

\begin{center}
APPENDIX\ II
\end{center}

In this appendix, we emphasize the similarity between the equation that gives
the energy shift as a function of the density and the elimination of $z$
between (\ref{1}) and (\ref{2}) that provides the equation of state.\ For
simplicity, we assume that no external potential acts on the particles, so
that translational invariance is satisfied.\ We call $\rho_{k}$ the diagonal
elements of $\rho_{I}$ :
\begin{equation}
\rho_{\mathbf{k}}=<\mathbf{k}\mid\rho_{I}\mid\mathbf{k}> \label{a10}%
\end{equation}
and $u_{1}(\mathbf{k})$ those of $U_{1}$:
\[
\mathbf{\;}u_{1}(\mathbf{k})=<\mathbf{k}\mid U_{1}\mid\mathbf{k}>=e^{-\beta
e_{k}}%
\]
where $e_{k}$ is the single particle kinetic energy. As for the diagonal
elements of $U_{2}$ in the momentum representation, we note them:
\[
u_{2}(\mathbf{k}_{1},\mathbf{k}_{2})=<\mathbf{k}_{1},\mathbf{k}_{2}\mid
U_{2}\mid\mathbf{k}_{1},\mathbf{k}_{2}>
\]
(for the moment we do not make the MIME approximation), $u_{3}(\mathbf{k}%
_{1},\mathbf{k}_{2},\mathbf{k}_{3})$ for those of $U_{3}$, etc. Equation
(\ref{a6}) then provides:
\begin{equation}%
\begin{array}
[c]{cl}%
\rho_{\mathbf{k}} & =u_{1}(\mathbf{k})\left[  z+z^{2}\sum_{\mathbf{k}%
^{^{\prime}}}u_{2}(\mathbf{k}_{1},\mathbf{k}_{2})\,u_{1}(\mathbf{k}^{^{\prime
}})\,+\right. \\
& \left.  +\frac{z^{3}}{2}\sum_{\mathbf{k}^{^{\prime}}}\sum_{\mathbf{k}%
^{^{\prime\prime}}}u_{3}(\mathbf{k}_{1},\mathbf{k}_{2},\mathbf{k}%
_{3})\,\,\,u_{1}(\mathbf{k}^{^{\prime}})u_{1}(\mathbf{k}^{^{\prime\prime}%
})\,+.......\right]
\end{array}
\label{a11}%
\end{equation}
Now, if we introduce the energy shift $\Delta(\mathbf{k})$ by the relation:
\begin{equation}
\rho_{\mathbf{k}}=z\,\,u_{1}(\mathbf{k})\times e^{-\beta\Delta(\mathbf{k})}
\label{a12}%
\end{equation}
we obtain:
\begin{equation}%
\begin{array}
[c]{cl}%
e^{-\beta\Delta(\mathbf{k})} & =1+z\sum_{\mathbf{k}^{^{\prime}}}%
u_{2}(\mathbf{k}_{1},\mathbf{k}_{2})\,\,u_{1}(\mathbf{k}^{^{\prime}})+\\
& +\frac{z^{2}}{2}\sum_{\mathbf{k}^{^{\prime}}}\sum_{\mathbf{k}^{^{\prime
\prime}}}u_{3}(\mathbf{k}_{1},\mathbf{k}_{2},\mathbf{k}_{3})\,\,u_{1}%
(\mathbf{k}^{^{\prime}})u_{1}(\mathbf{k}^{^{\prime\prime}})+...
\end{array}
\label{a13}%
\end{equation}
In this relation, we express the energy shift $\Delta(\mathbf{k})$ as a
function of a series in power of the populations $z\,u_{1}(\mathbf{k})$ of the
ideal gas; but if we write:
\begin{equation}%
\begin{array}
[c]{cl}%
e^{-\beta\Delta(\mathbf{k})} & =1+\sum_{\mathbf{k}^{^{\prime}}}u_{2}%
(\mathbf{k}_{1},\mathbf{k}_{2})\,\,e^{\beta\Delta(\mathbf{k}^{^{\prime}}%
)}\,\,\rho_{\mathbf{k}^{^{\prime}}}\\
& +\frac{1}{2}\sum_{\mathbf{k}^{^{\prime}}}\sum_{\mathbf{k}^{^{\prime\prime}}%
}u_{3}(\mathbf{k}_{1},\mathbf{k}_{2},\mathbf{k}_{3})\,\,\,e^{\beta\left[
\Delta(\mathbf{k}^{^{\prime}})+\Delta(\mathbf{k}^{^{\prime\prime}})\right]
}\,\rho_{\mathbf{k}^{^{\prime}}}\rho_{\mathbf{k}^{^{\prime\prime}}}+...
\end{array}
\label{a14}%
\end{equation}
we obtain another expression where only the actual populations $\rho
_{\mathbf{k}^{^{\prime}}}$ play a role, and from which the variable $z$ has
now disappeared. Finally, it is sufficient to take the logarithm of the right
hand side of this series to obtain the expression of the energy shift
$\Delta(\mathbf{k})$ as a function of the populations as well as all the other
energy shifts $\Delta(\mathbf{k}^{^{\prime}})$.

In the MIME approximation, the interactions constants $u_{2}$, $u_{3}$, etc.
factorize out of the sums, the energy shift becomes a constant $\Delta$ that
is independent of $\mathbf{k}$, and one merely obtains:
\begin{equation}
e^{-\beta\Delta}=1+u_{2}\,e^{\beta\Delta}\,N+\frac{u_{3}}{2}\,e^{2\beta\Delta
}\,N^{2}+... \label{a15}%
\end{equation}
where $N\equiv\left\langle N\right\rangle $ is the mean total number of
particles (the trace of $\rho_{I}$). This equation provides an implicit
equation between the energy shift $\Delta$ and the density $\rho
=N/\mathcal{V}$ (since $u_{2}$ is proportional to the inverse volume, $u_{3}$
to the square of this inverse volume, etc.) From this result, one can expand
the energy shift in powers of the density:
\begin{equation}
\Delta=u_{2}\ N+u_{3}^{^{\prime}}\,\,N^{2}+u_{4}^{^{\prime}}\,\,N^{3}+...
\label{a16}%
\end{equation}
where the new coefficients $u_{3}^{^{\prime}}$, $u_{4}^{^{\prime}}$, ... can
be calculated step by step by inserting this relation into (\ref{a15}); here
again, the first term in the right hand side of (\ref{a16}) corresponds to the
mean-field, and the following terms to density corrections.\ We note the
similarity between this result and the usual elimination of $z$ between the
two $z$-expansions of the pressure and the density, which provides the
equation of state, where only the actual density of the system appears. We
also remark that, in general, the energy shift has a contribution that is
linear in density, which can be called mean-field, but also higher order
density terms (density corrections to the mean-field).

\bigskip\bigskip

\begin{center}
APPENDIX III
\end{center}

We have defined in section \ref{tree} the reducible part $\overline{U}_{l}%
^{R}(1,2,...,l)$ of any Ursell operator $\overline{U}_{l}(1,2,...l)$ in terms
of ``trees'' made of chains of $\overline{U}_{2}$ operators.\ For instance,
the Ursell operator of order $l=3$ then becomes the sum of a tree-reducible
part and of an irreducible part:%
\begin{equation}
\overline{U}_{3}(1,2,3)=\overline{U}_{3}^{R}(1,2,3)\,+\,\overline{U}%
_{3}^{Irr.}(1,2,3) \label{app1}%
\end{equation}
In this appendix, we sketch how the notion of reducibility could be
generalized: the irreducible part $\overline{U}_{3}^{Irr.}\,$could be used,
exactly as $\overline{U}_{2}$, as a starting point to define a ``$\overline
{U}_{3}$-reducible'' part of all operators $\overline{U}_{l}$ with $l\geq4$ in
terms of products of $\overline{U}_{2}$'s and $\overline{U}_{3}^{Irr.}$'s.%
%TCIMACRO{\FRAME{ftbpFU}{1.5999in}{1.5489in}{0pt}{\Qcb{An example of a diagram
%including irreducible parts $\overline{U}_{3}^{Irr.}$ of the three particle
%Ursell operator $U_{3}$, symbolized by triangles, in addition to straight
%lines representing the $\overline{U}_{2}$'s.\ Such diagrams contribute to a
%tern in the energy shift which is quadratic in $\rho_{I}$.}}{\Qlb{figure7}%
%}{figure7.eps}{\special{ language "Scientific Word";  type "GRAPHIC";
%width 1.5999in;  height 1.5489in;  depth 0pt;  original-width 3.0995in;
%original-height 3in;  cropleft "0";  croptop "1";  cropright "1";
%cropbottom "0";  filename 'figure7.eps';file-properties "XNPEU";}} }%
%BeginExpansion
\begin{figure}
[ptb]
\begin{center}
\includegraphics[
height=1.5489in,
width=1.5999in
]%
{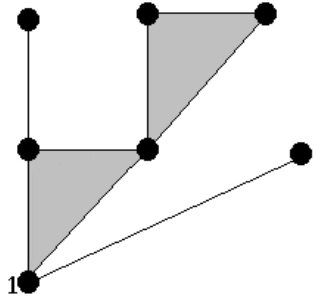}%
\caption{An example of a diagram including irreducible parts $\overline{U}%
_{3}^{Irr.}$ of the three particle Ursell operator $U_{3}$, symbolized by
triangles, in addition to straight lines representing the $\overline{U}_{2}%
$'s.\ Such diagrams contribute to a tern in the energy shift which is
quadratic in $\rho_{I}$.}%
\label{figure7}%
\end{center}
\end{figure}
%EndExpansion

The basic idea is to build trees, not only with lines that represent
$\overline{U}_{2}$'s, but also with triangles that represent $\overline{U}%
_{3}^{Irr.}$'s.\ Figure \ref{figure7} shows an example of such a tree: at each
branching point, one may now connect either lines, or one corner of a
triangle; in the latter case, the two other corners can be used as starting
points in order to extend the branch further, with any product of
$\overline{U}_{2}$'s and $\overline{U}_{3}^{Irr.}$'s. One difference is that,
because each $\overline{U}_{3}^{Irr.}$, once added, introduces two new
possible branching points (instead of a single one for a $\overline{U}_{2}$),
the question arises as to which one is represented first in the clockwise
order; it is easy to see that this introduces a $1/2$ factor per $\overline
{U}_{3}^{Irr.}$ contained in the diagram into the corresponding
weight.\ Otherwise, not much is changed, provided of course all branching
factors (including those resulting from $\overline{U}_{3}^{Irr.}$'s) are
included in the weight; in other words, in formula (\ref{13quater}), a factor
$(1/2)^{n_{3}}$ is added (where $n_{3}$ is the number of $\overline{U}%
_{3}^{Irr.}$'s), and the ramification factors $r_{i}$ now include the effect
of $\overline{U}_{3}^{Irr.}$'s.\ This being done, the property of self
similarity appears again in the series, and the essence of the reasoning still
holds with relatively minor changes.

Finally, one is led to the introduction of an exponential:%
\begin{equation}
e^{-\beta\Delta(1)} \label{app2}%
\end{equation}
where $\Delta(1)$ would now be given by the sum:%
\begin{equation}
-\beta\Delta(1)=Tr_{2}\left\{  \overline{U}_{2}(1,2)\rho_{I}(2)\right\}
+Tr_{2,3}\left\{  \frac{1}{2!}\overline{U}_{3}^{Irr.}(1,2,3)\rho_{I}%
(2)\rho_{I}(3)\right\}  \label{app3}%
\end{equation}
In other words, the energy shift would no longer be proportional to the single
particle density operator,\ but would contain a term that is quadratic in
$\rho_{I}$. Similarly, one can expect that the irreducible part $\overline
{U}_{4}^{Irr.}$ of $\overline{U}_{4}$ would lead to a contribution to the
energy shift that is cubic in $\rho_{I}$, and so on. We have not yet performed
these calculations.


\begin{thebibliography}{9}                                                                                                %
\bibitem {Green-1}L.P.\ Kadanoff \& G.\ Baym, ``Quantum Statistical
Mechanics'', Benjamin (1962).

\bibitem {Green-2}A.A.\ Abrikosov, L.P. Gorkov \& I.E. Dzyaloshinski,
``Methods of Quantum Field Theory in Statistical Physics'', Dover (1963).

\bibitem {Green-3}A.L.\ Fetter \& J.D.\ Walecka, ``Quantum Theory of
Many-Particle Systems'', McGraw Hill (1971).

\bibitem {Ursell-1}P.\ Gr\"{u}ter \& F.\ Lalo\"{e}, J.\ Physique \textbf{5},
181 (1995).

\bibitem {Ursell-1-bis}P.\ Gr\"{u}ter \& F.\ Lalo\"{e}, J.\ Physique
\textbf{5}, 1255 (1995).

\bibitem {Ursell-2}M.\ Holzmann, P.\ Gr\"{u}ter \& F.\ Lalo\"{e},
Eur.\ Phys.\ J.\ \textbf{B10}, 739 (1999).

\bibitem {Mayer}J.E. Mayer \& M. Goeppert Mayer, ``Statistical Mechanics''
2$^{nd}$edition, Wiley (1975).

\bibitem {Ursell original}H.D. Ursell, Proc. Camb. Phil. Soc. \textbf{23}, 685 (1927).

\bibitem {Uhlenbeck}G.E. Uhlenbeck \& G.W. Ford, ``Lectures in Statistical
Mechanics'', American Mathematical Society (1962).

\bibitem {Hansen}J.P. Hansen \& I.R. McDonald, ``Theory of Simple Liquids''
2$^{nd}$edition, Academic Press (1986), chap.4.

\bibitem {Mayer bis}J.E. Mayer \& M. Goeppert Mayer loc. cit. p.258.

\bibitem {Hansen bis}J.P. Hansen \& I.R. McDonald loc. cit. p.91.

\bibitem {Morita}T. Morita \& K. Hiroike, Prog. Th. Phys. \textbf{25}, 537 (1961).

\bibitem {Dominicis}C. de Dominicis, J.\ Math.\ Phys.\ \textbf{3}, 983 (1962)
and \textbf{4}, 255 (1963).

\bibitem {Van-Kampen}N.G.\ Van Kampen, Physica \textbf{27}, 783 (1961).

\bibitem {Mullin}W.J.\ Mullin, Am.\ J.\ Physics \textbf{40}, 1473 (1972).

\bibitem {Lee}T.D. Lee \& C.N. Yang, Phys. Rev. \textbf{87}, 404 (1952).

\bibitem {Markus}M.\ Holzmann, PhD\ thesis, Paris (2000).

\bibitem {Ursell-1-ter}P.\ Gr\"{u}ter, F.\ Lalo\"{e}, A.E.\ Meyerovich \&
W.J.\ Mullin, J.\ Physique \textbf{7}, 485 (1997).

\bibitem {BBHLV}G.\ Baym, J.P.\ Blaizot, M.\ Holzmann, F.\ Lalo\"{e},
D.\ Vautherin, Phys. Rev. Lett. \textbf{83}, 1703 (1999).

\bibitem {Kahn}B. Kahn \& G.E. Uhlenbeck, Physica \textbf{5}, 399 (1938).
\end{thebibliography}
\end{document}